%% file: Dash2.tex
\documentclass[useAMS,usenatbib,fleqn]{mn2e}

\usepackage{hyperref}

\usepackage{graphicx}
\usepackage{amsmath}
\usepackage{amssymb}
\usepackage{dcolumn}

\newcommand{\mcc}[1]{\multicolumn{1}{c}{#1}}

\newcommand{\rs}{\ensuremath{{R_{\star}}}}

\newcommand{\lx}{\ensuremath{L_\mathrm{X}}}
\newcommand{\lbol}{\ensuremath{L_\mathrm{Bol}}}

\title[X-ray emission from NGC\,1624-2]{X-ray emission from the giant magnetosphere of the magnetic O-type star NGC\,1624-2%
}

\author[Petit et al.]{V. Petit$^{1}$\thanks{E-mail: VPetit@fit.edu},
D. H. Cohen$^2$,
G. A. Wade$^3$,
Y. Naz\'e$^4$\thanks{FRS-FNRS research associate},
S. P. Owocki$^5$,\newauthor
J. O. Sundqvist$^5$,
A. ud-Doula$^6$, 
A. Fullerton$^7$,
M. Leutenegger$^{8,9}$,
M. Gagn\'e$^{10}$
\\
$^{1}$ Dept. of Physics \& Space Sciences, Florida Institute of Technology, Melbourne, FL 32904, USA\\
$^{2}$ Dept. of Physics \& Astronomy, Swarthmore College, Swarthmore, PA 19081, USA  \\
$^{3}$ Dept. of Physics, Royal Military College of Canada, PO Box 17000, Stn Forces, Kingston, Ontario K7K 7B4, Canada \\
$^{4}$ GAPHE, Universit\'e de Li\`ege, Quartier Agora, All\'ee du 6 Ao\^ut 19c, Bat. B5C, B-4000 Li\`ege, Belgium \\
$^{5}$ Dept. of Physics \& Astronomy, University of Delaware, Bartol Research Institute, Newark, Delaware 19716, USA \\
$^{6}$ Penn State Worthington Scranton, Dunmore, PA 18512, USA \\
$^7$Space Telescope Science Institute, 3700 San Martin Dr., Baltimore, MD 21218, USA\\
$^8$NASA/Goddard Space Flight Center, Code 662, Greenbelt, MD 20771, USA \\
$^9$CRESST and University of Maryland, Baltimore County, Baltimore, MD 21250, USA\\
$^{10}$Department of Geology \& Astronomy, West Chester University, West Chester, PA 19383, USA}

\begin{document}
\include{aas_macros}

\date{
Accepted 2015 July 28.  Received 2015 July 6; in original form 2015 May 15}
\pagerange{\pageref{firstpage}--\pageref{lastpage}} \pubyear{2014}
\maketitle
\label{firstpage}

\begin{abstract}

We observed NGC\,1624-2, the O-type star with the largest known magnetic field ($B_p\sim{20}$\,kG), in X-rays with the ACIS-S camera onboard the Chandra X-ray Observatory. Our two observations were obtained at the minimum and maximum of the periodic H$\alpha$ emission cycle, corresponding to the rotational phases where the magnetic field is the closest to equator-on and pole-on, respectively. With these observations, we aim to characterise the star's magnetosphere via the X-ray emission produced by magnetically confined wind shocks.
Our main findings are:
(i) The observed spectrum of NGC\,1624-2 is hard, similar to the magnetic O-type star $\theta^1$\,Ori\,C, with only a few photons detected below 0.8\,keV. The emergent X-ray flux is 30\% lower at the H$\alpha$ minimum phase.
(ii) Our modelling indicated that this seemingly hard spectrum is in fact a consequence of relatively soft intrinsic emission, similar to other magnetic Of?p stars, combined with a large amount of local absorption ($\sim$1-3$\times10^{22}$\,cm$^{-2}$). This combination is necessary to reproduce both the prominent Mg and Si spectral features, and the lack of flux at low energies. 
NGC\,1624-2 is intrinsically luminous in X-rays ($\log L^{\mathrm{em}}_\mathrm{X}\sim$33.4) but 70-95\% of the X-ray emission produced by magnetically confined wind shocks is absorbed before it escapes the magnetosphere ($\log L^{\mathrm{ISMcor}}_\mathrm{X}\sim$32.5).  
(iii) 
The high X-ray luminosity, its variation with stellar rotation, and its large attenuation are all consistent with a large dynamical magnetosphere with magnetically confined wind shocks.

\end{abstract}


\defcitealias{2012MNRAS.425.1278W}{W12a}

\section{Introduction}

\begin{figure}
	\includegraphics[width=0.5\textwidth]{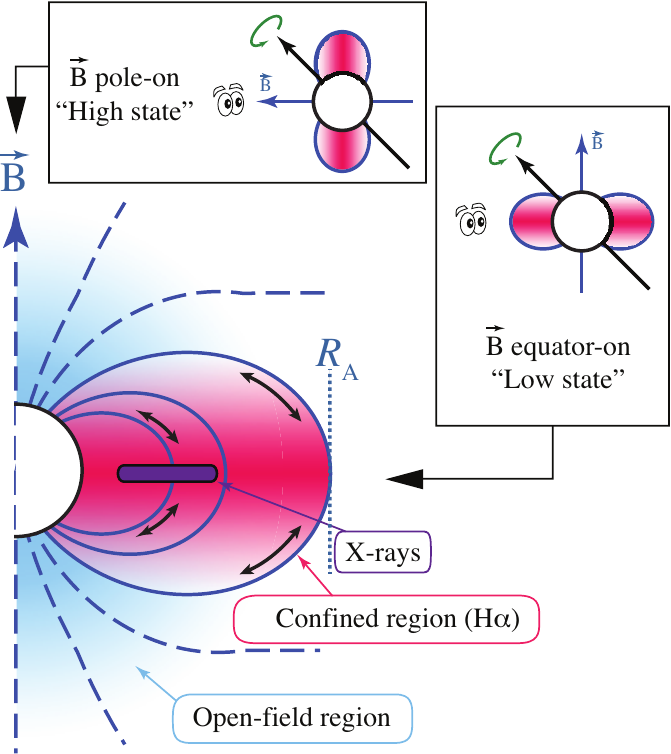}
	\caption{\label{fig:cartoon} Schematic of a magnetic massive star dynamical magnetosphere \citep[e.g.][]{2012MNRAS.423L..21S,2013MNRAS.429..398P}. 	Solid blue lines indicate regions below the last closed magnetic loop that confine the wind, located near the Alfv\'en radius $R_A$. Most of the H$\alpha$ emission originates here. Dashed lines indicate regions where the momentum of the wind results in open field lines. The bulk of the X-rays are produced in the region indicated in purple; see  \S\ref{sec:disc}.
The insets illustrate the view of an observer as the star's rotation changes the orientation of the magnetosphere. It is important to note that due to the long rotation periods of magnetic O-type stars, the dynamical effects of rotation on the magnetospheric structure are negligible \citep{2008MNRAS.385...97U}.} %
\end{figure}

A small fraction of O stars have large-scale, stable surface magnetic fields, as revealed in only the past decade by spectropolarimetric measurements of the Zeeman effect \citep[e.g.][]{2006MNRAS.365L...6D, 2010MNRAS.407.1423M,2012MNRAS.419.2459W}.
Even before these direct detections, an oblique dipole field was suspected in the prototype magnetic O star, $\theta^1$\,Ori\,C, notably due to its luminous, rotationally modulated X-ray emission \citep{1996A&A...312..539S,1997ApJ...478L..87G}. Indeed, strong and rotationally modulated X-ray emission due to a combination of occultation and local absorption was predicted to be the hallmark of the wind-fed dynamical magnetospheres of magnetic O stars with dipole fields misaligned with their rotational axis \citep[e.g.][]{1997A&A...323..121B,2005ApJ...628..986G,2014MNRAS.441.3600U}. 
The X-rays are due to the radiative cooling of plasma that is shock-heated when wind-flows from opposite hemispheres collide near the magnetic equatorial plane (see Fig.\,\ref{fig:cartoon}). Measurements of this X-ray emission are therefore a primary diagnostic of the physical conditions in, and processes that control, the magnetospheres of O stars.  

The rotational modulation of the X-ray emission in $\theta^1$\,Ori\,C is due to occultation of the post shock material by the stellar disk during equator-on viewing phases, with no conclusive evidence in phase-resolved high-resolution \textit{Chandra} spectra for any large absorption of X-rays by the magnetosphere, despite the fact that absorption by cooled post-shock plasma was expected \citep{1997ApJ...485L..29B,2005ApJ...628..986G,2002MNRAS.333...55D}. 
The surface field strength of a little over 1\,kG implies a maximum closure radius of magnetic field lines\footnote{located near the Alfv\'en radius $R_A$, see equations 40 and 41 of \citet{2014MNRAS.441.3600U}.} of only slightly more than 2$\rs$.
The modest size of this star's magnetosphere therefore leads to relatively low absorption column densities.

Since the direct detection of the magnetic field on $\theta^1$\,Ori\,C \citep{2002MNRAS.333...55D}, significant magnetic fields have been measured on $\sim$10 other O stars \citep[see][]{2013MNRAS.429..398P,2014Msngr.157...27M}. 
Significantly, most of these stars are classified as Of?p \citep{2010ApJ...711L.143W}, though the reason for this connection between spectral morphology and magnetism is not yet fully understood. All magnetic O stars share physical and magnetic properties with $\theta^1$\,Ori\,C.
As discussed by \citet{2013MNRAS.429..398P} all but one of the magnetic O-type stars are slow rotators, consistent with strong magnetic braking associated with their powerful stellar winds \citep{2009MNRAS.392.1022U}. Their observed optical magnetospheric properties are well explained within the dynamical magnetosphere framework: with negligible rotational support, the plasma environment in the magnetosphere should be quite dynamic, with trapped plasma falling back down onto the star in not much longer than a free-fall time, as confirmed by numerical MHD simulations \citep{2013MNRAS.428.2723U}.  

Magnetic O stars are generally an order of magnitude more luminous in X-rays than non-magnetic O stars of the same spectral subtype, and their X-rays are somewhat harder, although the X-ray emission from other Of?p stars is generally not as hard as that of $\theta^1$\,Ori\,C  \citep{2014ApJS..215...10N}. Given the similarity in physical and magnetic properties among the magnetic O-type stars, this difference in spectral X-ray properties has yet to be understood. 

The star that is the subject of this paper, the slowly rotating O-star NGC\,1624-2, has a field strength more than an order of magnitude larger than any other known magnetic O-stars, at nearly 20\,kG  \citep[][hereafter \citetalias{2012MNRAS.425.1278W}]{2012MNRAS.425.1278W}. 
Given that field strength, and a similar expected wind mass-loss rate as $\theta^1$\,Ori\,C and the Of?p stars, it should have a significantly larger magnetic closure radius and thus a significantly larger dynamical magnetosphere, extending out to $\sim 11~R_\star$.
In order to explore the X-ray generation and also the possible magnetospheric absorption of X-rays in this star's giant magnetosphere, we have obtained two \textit{Chandra} ACIS observations of NGC\,1624-2 at rotational phases that correspond to two viewing angles, respectively when the magnetic pole and equator are closest to the line of sight. 

In this paper we present detailed analysis of the X-ray spectral properties at both phases, including modelling of any possible local absorption by partially ionised plasma in the magnetosphere. This allows us to draw conclusions about the physical properties of the magnetically confined wind shocks, the wind feeding rate and heating efficiency, the post-shock cooling plasma, and the spatial distribution of magnetospheric material. 
In addition to this empirical spectral modelling, we present a comparison with recent MHD models of X-ray emission from magnetospheres \citep{2014MNRAS.441.3600U}. Also preliminary results from the post-processing of MHD simulations of a generic O star magnetosphere provide information about the expected viewing-angle dependence of the X-ray absorption column density (Owocki et al. in preparation). 
The goal of this paper is to constrain models and extend our understanding of the dynamical magnetospheres of O stars via observations of the most extreme example currently known.

\section{Stellar properties of NGC\,1624-2}

NGC\,1624-2 is an O7\,f?cp star that was recently singled out as the most extreme example of the Of?p spectroscopic category \citep{2010ApJ...711L.143W}. 
Soon after, a $\sim$20\,kG field was derived by the MiMeS Collaboration \citepalias{2012MNRAS.425.1278W}.
Although very strong magnetic fields have been measured for a handful of lower mass ApBp stars \citep[e.g. 34\,kG Babcock's star; see][]{1997A&AS..123..353M}, NGC\,1624-2 has by far the strongest magnetic field ever measured on an O-type star: all the others have $B_\mathrm{pole}<2.5$\,kG.  
According to the properties derived by \citetalias[][summarised here in Table\,\ref{tab:prop}]{2012MNRAS.425.1278W}, NGC\,1624-2's magnetosphere is estimated to extend to $R_\mathrm{A}\approx11R_\star$ hence trapping 95\% of the outflowing wind, much more than other magnetic O stars which have $R_\mathrm{A}$ of just a few stellar radii. 

\citetalias{2012MNRAS.425.1278W} determined a rotational period of 157.99\,d from the observed spectral variations. This slow rotation is consistent with the circular polarisation signal variation, and with the narrow photospheric line profiles. As pointed out by \citet{2013MNRAS.433.2497S}, the width of the spectral lines, which exhibit Zeeman splitting, is completely accounted for by Zeeman broadening, and does not show any signs of the large macroturbulent broadening generally observed in both normal and magnetic O-type stars. 
This suggests that NGC\,1624-2 is the only known magnetic O-type star with a field strong enough to stabilise its deep photosphere, providing the first empirical constraints on the connection between the sub-surface convection zone present in models of hot, massive stars \citep{2009A&A...499..279C}, and macro-turbulent line broadening. 

The dense magnetospheres of magnetic O stars produce emission in optical recombination lines, such as the Balmer lines, which is modulated according to the rotational period. This variation is due to changes in the projected distribution of optically thick magnetospheric material \citep[see Fig.\,\ref{fig:cartoon};][]{2012MNRAS.423L..21S}.
Although these slowly rotating magnetospheres are quite dynamic in nature, their volume-integrated properties are quite stable \citep{2007MNRAS.381..433H,2013MNRAS.428.2723U} from one rotation cycle to another.

Although the detailed magnetic geometry of NGC\,1624-2 has yet to be established, the single-wave, sinusoidal nature of the spectral line variation implies that we only view one magnetic hemisphere of a mainly dipolar field during the rotation of the star \citepalias{2012MNRAS.425.1278W}. However, the large variation in the H$\alpha$ emission (Fig.\,\ref{fig:ha}) between a ``high state'', when the magnetic pole is most closely aligned with the observer's line of sight, and a ``low state'', when the magnetic equator is closest to the line of sight \citep[e.g.][]{2012MNRAS.423L..21S} indicates a large change in our viewing angle (see Fig.\,\ref{fig:cartoon}).

NGC\,1624 was observed for 10\,ks with the \textit{Chandra} X-ray Observatory in 2006 (ObsID 7473, PI Gordon).
Even though NGC\,1624-2 was detected with only 38 counts, only one count has an energy below 1\,keV indicating rather hard emission. According to the analysis of \citetalias{2012MNRAS.425.1278W}, this spectrum could be reproduced with a hot ($\sim2.3$\,keV) plasma, similar to that of $\theta^1$\,Ori\,C. It could also be reproduced with a cooler ($\sim0.7$\,keV) plasma, similar to that of the other Of?p stars, that would be more heavily absorbed ($N_\mathrm{H}\sim2\times10^{22}$\,cm$^{-2}$). 
We note that even during the low state, the H$\alpha$ emission of NGC\,1624-2 is much stronger than that of any other known magnetic O-type star \citepalias[][see their Fig.\,11]{2012MNRAS.425.1278W}. This suggests that the magnetosphere must indeed be very large, and higher S/N X-ray observations are the perfect tool to verify this hypothesis through the presence of X-ray absorption.

\begin{table}
\centering
\caption{\label{tab:prop}Summary of adopted stellar, wind, magnetic, and magnetospheric properties of NGC 1624-2 from \citetalias{2012MNRAS.425.1278W}. 
The stellar and magnetic parameters were derived from optical spectropolarimetry. The unperturbed wind mass-loss rate $\dot M$ and terminal velocity $v_\infty$ were calculated based on the theoretical prescriptions of \citep{2001A&A...369..574V}. 
The wind magnetic confinement Alfven radius $R_{\rm Alf}$, and the Kepler co-rotation radius $R_{\rm Kep}$ are calculated as described by \citet{2013MNRAS.429..398P}.}
\begin{tabular}{ll}
\hline
Spectral type &  O6.5f?cp - O8f?cp            \\
$T_{\rm eff}$ (K) & 35 000 $\pm$ 2000 \\
log $g$ (cgs) & 4.0 $\pm$ 0.2     \\
R$_{\star}$ (R$_\odot$) & $10\pm 3$ \\
$\log (L_\star/L_\odot)$ & $5.10\pm 0.2$   \\
{$M_{\star}^{\rm evol}$ ($M_{\odot}$)} & {$28^{+7}_{-5}$} \\
{$\log \dot{M}_{B=0}$} (M$_{\odot}$\,yr$^{-1}$) & $-6.8$  \\
$v_{\infty, B=0}$ (km\,s$^{-1}$) & 2875  \\
$E(B-V)$ (mag) &$0.802\pm 0.02$ \\
$R_{\rm V}$ & $3.74\pm 0.1$ \\
$\log d$ (pc)  & 3.71 $\pm$ 0.1 \\
$P_\mathrm{rot}$ (d) & $157.99 \pm 0.94$ \\
$B_{\rm d}$ (kG) & $\sim 20$ \\
R$_{\rm Alf}$ ($R_*$) & $11$ \\
R$_{\rm Kep}$ ($R_*$) & $>40$ \\
\hline
\end{tabular}
\end{table}

\section{Observations}

\begin{figure}
	\includegraphics[width=0.5\textwidth]{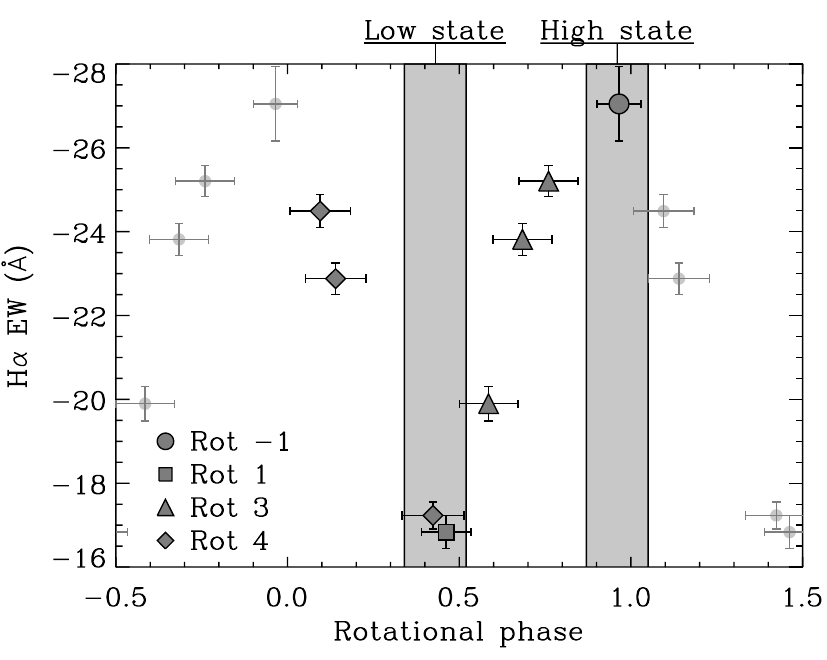}
	\caption{\label{fig:ha} Equivalent width of the H$\alpha$ line of NGC\,1624-2 phased with the ephemeris of \citetalias{2012MNRAS.425.1278W}. The symbol shapes represent the rotation number since the ephemeris $JD_0$. 
	The optical spectra were obtained with ESPaDOnS at CFHT over a period of time corresponding to nearly 3 rotations of the star, by the MiMeS Collaboration \citep{2010arXiv1009.3563W} and on PI time (Wade).
	The \textit{Chandra} observations, indicated by the grey shaded area, were obtained during rotation cycle no. 3.} 
\end{figure}

We obtained two observations of NGC\,1624-2 with the bare ACIS-S camera in VFaint mode aboard the \textit{Chandra} X-ray Observatory (PI: Petit). The ACIS-S camera allows for imaging spectroscopy combining moderate spectral resolution ($FWHM$$\simeq$150\,eV at 1.5\,keV) and sub-arcsec spatial resolution. 
According to the ephemeris of \citetalias{2012MNRAS.425.1278W} ($JD = [245\,5967.0 \pm 10] + [157.99 \pm 0.94] \times E$), the observations were obtained at rotational phase $0.43\pm 0.09$ (ObsID 14572) and $0.96\pm 0.09$ (ObsID 14571). These correspond roughly to the low state and the high state, respectively (see Fig\,\ref{fig:ha}).

\begin{figure*}
	\includegraphics[width=\textwidth]{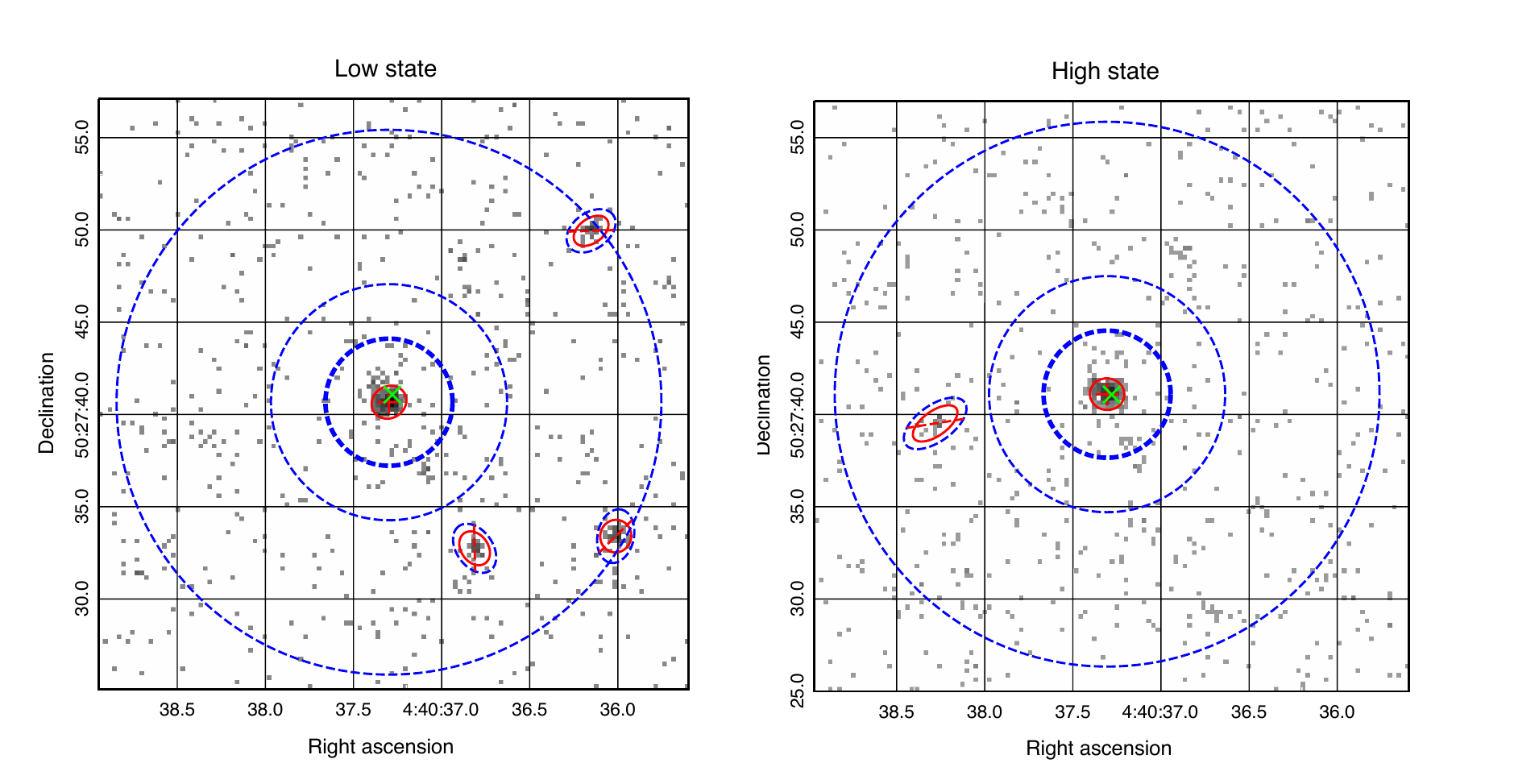}
	\caption{\label{fig:ds9} \textit{Chandra} ACIS-S images of NGC\,1624-2 (position marked by the green cross) during the H$\alpha$ low state (left) and high state (right). The sources detected by \textsc{celldetect}, including NGC\,1624-2, are indicated by solid red ellipses. The spectral extraction source (thick) and background (thin) regions are shown as dashed blue circles.  The other point sources circled in dashed blue were excluded from the background estimate.  } 
\end{figure*}

The data were reprocessed according to standard reduction procedures\footnote{The {\sc ciao} analysis threads are available at\\ \href{http://cxc.harvard.edu/ciao/threads/index.html}{http://cxc.harvard.edu/ciao/threads/index.html}} with \textsc{ciao} version 4.5. 
Using the \textsc{celldetect} tool, an X-ray source is detected at the position of NGC 1624-2 in both observations (Fig.\,\ref{fig:ds9}).
The pile-up fraction, evaluated from maps of count rate per frame, is less than one percent for both observations. Table~\ref{tab:aprates} compiles the aperture photometry calculated with the \textsc{aprates} tool. The net photon flux is 40 percent higher during the H$\alpha$ emission high state than at the low state.

\begin{table}
\caption{\label{tab:aprates}Observation details and aperture photometry results.}
\renewcommand{\arraystretch}{1.5}
\input{Table2}
\end{table}

\section{Timing analysis}

\begin{figure*}
	\includegraphics[width=0.5\textwidth]{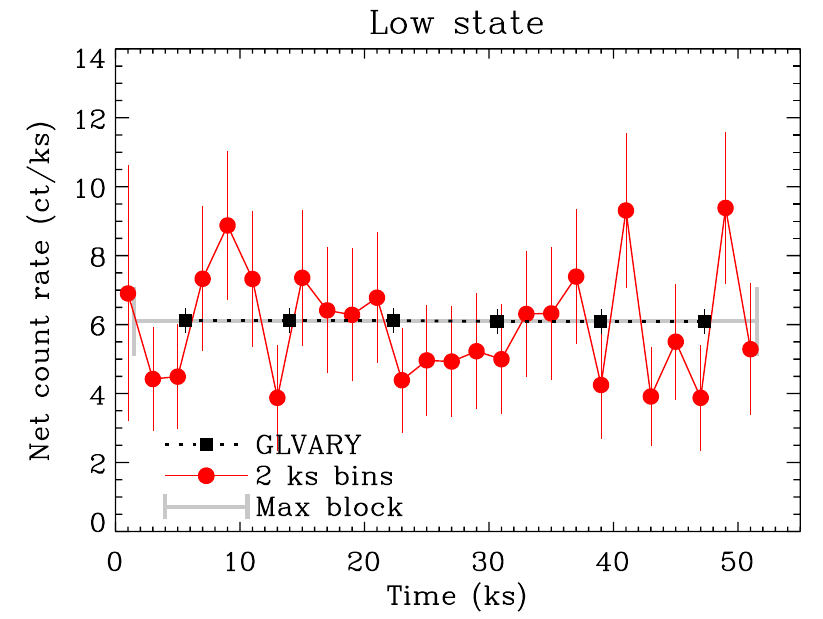}\includegraphics[width=0.5\textwidth]{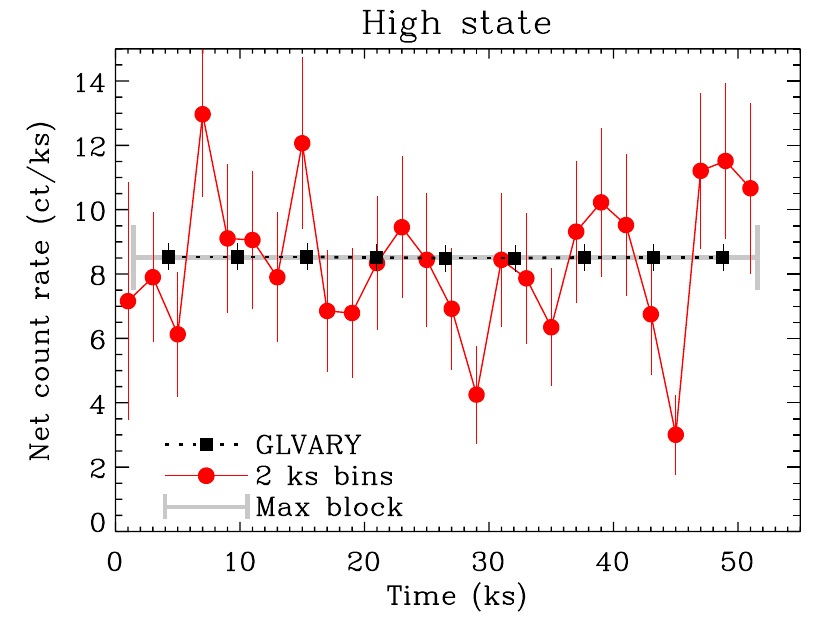}
	\caption{\label{fig:light_curve}Light curves between 0.3 and 10 keV, binned by 2\,ks (red circles). The probability-weighted light curves from the Gregory-Loredo variability algorithm implemented in the \textsc{glvary} routine are shown with black squares. The maximum-likelihood block algorithm resulted in a single block for each observation, shown by the grey bars.}
\end{figure*}

For both observations, we produced light curves to search for variability on a time scale comparable or less than the 50\,ks exposure. Such a timescale corresponds to only 0.4 percent of a rotation cycle, but is comparable to the wind flow timescale. The broad-band background-subtracted light curves, binned to an arbitrary 2\,ks, are shown in red circles in Fig.~\ref{fig:light_curve}. 
We applied the Gregory-Loredo variability algorithm to the source event list, as implemented in the \textsc{ciao-4.5} routine \textsc{glvary}. No evidence of short term variability is found and the resulting probability weighted light curves are shown as black squares. 
We also applied a maximum-likelihood block algorithm that divides the event list into blocks of constant count rate \citep{2005ApJS..160..423W}. Each epoch is reproduced by a single block (grey bar), which also indicates a lack of short-term variability. A comparison of the two epochs however confirms the statistically significant variability on the longer rotational timescale.

\section{Spectral analysis}

\begin{figure}
	\includegraphics[width=0.47\textwidth]{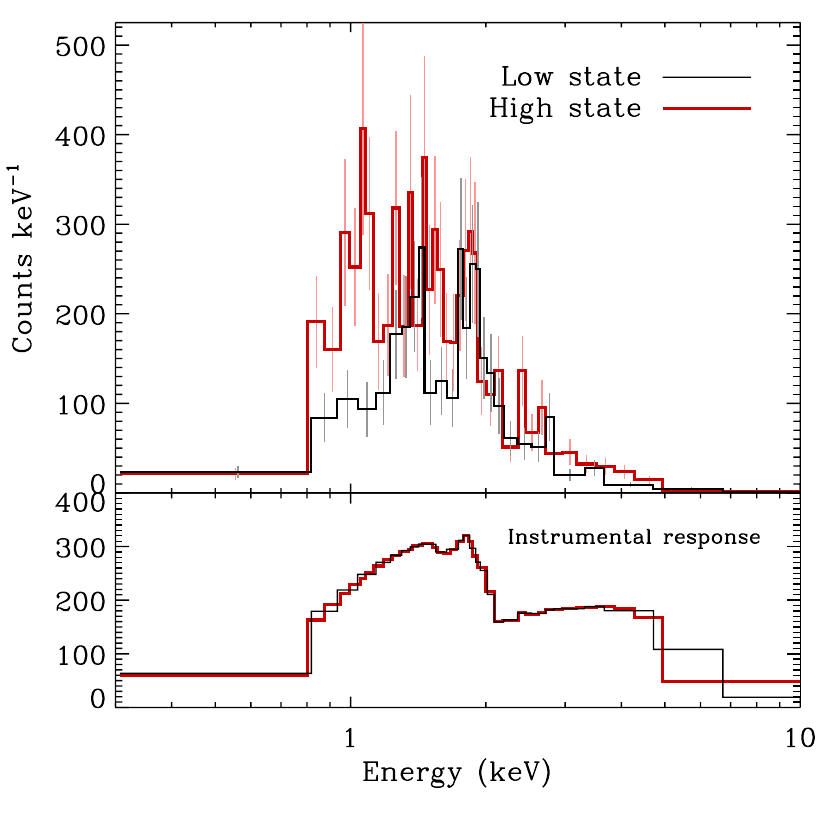}
	\caption{\label{fig:comp_spec}ACIS-S spectra of NGC\,1624-2 during the low state (thin black) and the high state  (thick red). The bottom panel shows a representation of the instrumental response, i.e. the spectra that would be observed if the emission model was flat. The small differences between the two epochs are caused by slight variations in response and adaptive signal-to-noise binning.}
\end{figure}

We extracted the spectra of NGC\,1624-2 and associated instrumental responses using the standard \textsc{specextract} procedure. We grouped the spectral channels to obtain a signal-to-noise ratio of 3 in each bin, allowing the use of $\chi^2$ statistics in our model fitting.
Fig.~\ref{fig:comp_spec} shows the resulting spectra (top panel) for the low state (thin black) and high state (thick red) observations. 
The bottom panel shows a representation of the spectral response functions for both observations. They are practically identical, allowing meaningful direct visual comparison of the spectra. 
In both observations, most of the photons have energies higher than 0.8\,keV. The high state observation shows an excess of soft ($\sim$1\,keV) photons. Prominent spectral features, similar in both epochs, are seen between 1 and 2\,keV.

We modeled the spectra with \textsc{xspec} 12.7 \citep{1996ASPC..101...17A} with an optically thin plasma model from the Astrophysical Plasma Emission Code \citep[APEC with \textsc{atomdb} 2.0.1,][]{2001ApJ...556L..91S,2012ApJ...756..128F}, which assumes collisional-radiative equilibrium (`coronal emission').
Given the lack of detailed abundance analysis in the optical  \citepalias{2012MNRAS.425.1278W}, we here adopted solar abundance values \citep{2009ARA&A..47..481A}.

X-rays in O-type stars are generally produced by shock-heated wind material that mostly cools by radiative emission.
If the wind material is continuously heated and cooled, a time-averaged description of the wind should contain a continuous distribution of plasma temperatures \citep[e.g.][]{2004ApJ...611..434A}. 
We therefore mimic such a distribution by using 4 plasma temperature components (\textit{T}$_1$+\textit{T}$_2$+\textit{T}$_3$+\textit{T}$_4$) with fixed $kT$ of 0.2, 0.6, 1.0, and 4.0 keV, corresponding loosely to the peak emissivity of dominant spectral features likely to be present in the ACIS-S bandpass (H-like and He-like atoms of Ne, Mg, Si, S, and Fe). 
Given that these emissivity functions generally extend over a broad range of temperatures, we believe this approach is sufficient to model the characteristics of the emission spectrum while minimising the number of fitting parameters \citep[see discussion in][]{2014ApJS..215...10N}. It is important to note that other representations of the post-shock plasma (e.g. different combinations of temperatures and their associated emission measures) can result in a similar emission spectrum, especially at the relatively low resolution of ACIS-S. However the goal of this analysis is to investigate the global properties of the phase-dependent absorption, and not the detailed distribution of emission measure.

We attenuate the total emission model spectrum described above with the cold ISM absorption model \textit{tbabs} [\textit{ColdAbs}$\times$(4\textit{T})] of \citet{2000ApJ...542..914W}. We adopt an ISM column density $N_\mathrm{H}^\mathrm{ISM}=0.48\times10^{22}$\,cm$^{-2}$, determined from $E(B-V)=0.804$, $R_\mathrm{V}=3.74$ \citepalias{2012MNRAS.425.1278W}, which results in $A_\mathrm{V}=3.01$ \citep{2009MNRAS.400.2050G}. 

In addition to this ``ISM only'' model, we consider the possibility of extra local circumstellar absorption (``LOC'') in two ways. 
First, we consider an additional cold absorption (``Cold LOC'') with the same absorption model as the ISM [\textit{ColdAbs}$\times$(\textit{ColdAbs}$\times$(4\textit{T}))]. 
Second, we consider the possibility that this additional absorber is composed of warm wind material at near the $\sim$35kK effective temperature of the star (``Warm LOC''; [\textit{ColdAbs}$\times$(\textit{WarmAbs}$\times$(4\textit{T}))]). 
Because H and He are mostly ionized at this temperature, the wavelength-dependent opacity will be different than for a cold absorber model, transmitting more flux at lower energies. 
We therefore use the model \textit{slabtabs} provided in the contributed \textsc{xspec} package \textsc{windprofile}\footnote{\href{http://heasarc.nasa.gov/xanadu/xspec/models/windprof.html}{http://heasarc.nasa.gov/xanadu/xspec/models/windprof.html}}. \textit{Slabtabs} models absorption by a simple slab of ionized material, with elemental abundance and ionization balance chosen for the application, and using opacities from \citet{1995A&AS..109..125V} as implemented by \citet{2010ApJ...719.1767L}.
We use their opacity model calculated for a generic O-type star wind with solar abundances (see their Fig.\,6) 

We consider a simplified geometric model where the absorber is located between the emission site and the observer. In this case, the attenuation goes as $\mathrm{e}^{-\tau_\lambda}$, where $\tau_\lambda$ is the wavelength-dependent optical depth.
It is likely that the sites of X-ray emission are embedded within the warm wind material. 
Such geometric models exist for spherically symmetric non-magnetic winds \citep[e.g. \textit{windtabs}][]{2010ApJ...719.1767L}, however they might not be readily applicable here given that a magnetosphere is not expected to be spherically symmetric.

We proceed to model the observations using two different approaches, which correspond to the results presented in the following two subsections:

\begin{enumerate}

\item 

In the simplest case (``individual models''), we let all model parameters (column density $nH^{\mathrm{LOC}}$ and the distance-weighted emission measure $EM_i$ associated with each temperature component) vary independently from one observation to the other. 
More specifically, this assumes that the plasma temperature distribution (the 4\textit{T} components) can differ from one epoch to another. This could be due to intrinsic changes in the temperature distribution with time, or by occultation of spatially stratified temperature distribution (see below).

\item 

The spectral variations of magnetic stars are usually understood to be due to a change in the observer's viewing angle as the tilted magnetosphere co-rotates with the stellar surface.
Although these magnetospheres are dynamic in nature, there are many indications that their volume-integrated properties are quite stable, namely based on numerical simulations \citep{2013MNRAS.428.2723U}, on observational diagnostics \citep{2007MNRAS.381..433H,2013ApJ...769...33T}, and on the lack of short term variability observed here.
With this picture in mind, we therefore do not expect the intrinsic X-ray emissivity of the magnetosphere itself to change significantly in time, but rather its sky-projected structure to modulate with the rotational phase. 
We thus also attempt to simultaneously model both epochs with a common 4\textit{T} emission model (``Joint models''). 
We allow for a change in the overall flux of the emission model spectrum during the low state [\textit{Constant}$\times$(4\textit{T})] to account for a possible occultation of the circumstellar X-ray emitting plasma by the stellar disk when the magnetic field is seen near equator-on \citep{2005ApJ...628..986G}. 
Since the column density depends on the ray's path, and therefore on the viewing angle, the local absorption is also allowed to differ between the two epochs. 
This, however, makes the simplifying assumption that there is no strong spatial gradient of plasma temperature inside the magnetosphere, which would require more precise models to constrain. 
In summary, the seven fitted parameters for the joint model are the column densities for the low and high states, the distance-weighted emission measure associated with each temperature component, and a flux reduction factor $f$ for the low state.

\end{enumerate}

\subsection{Results from individual models}

\begin{figure*}
	\includegraphics[width=0.5\textwidth]{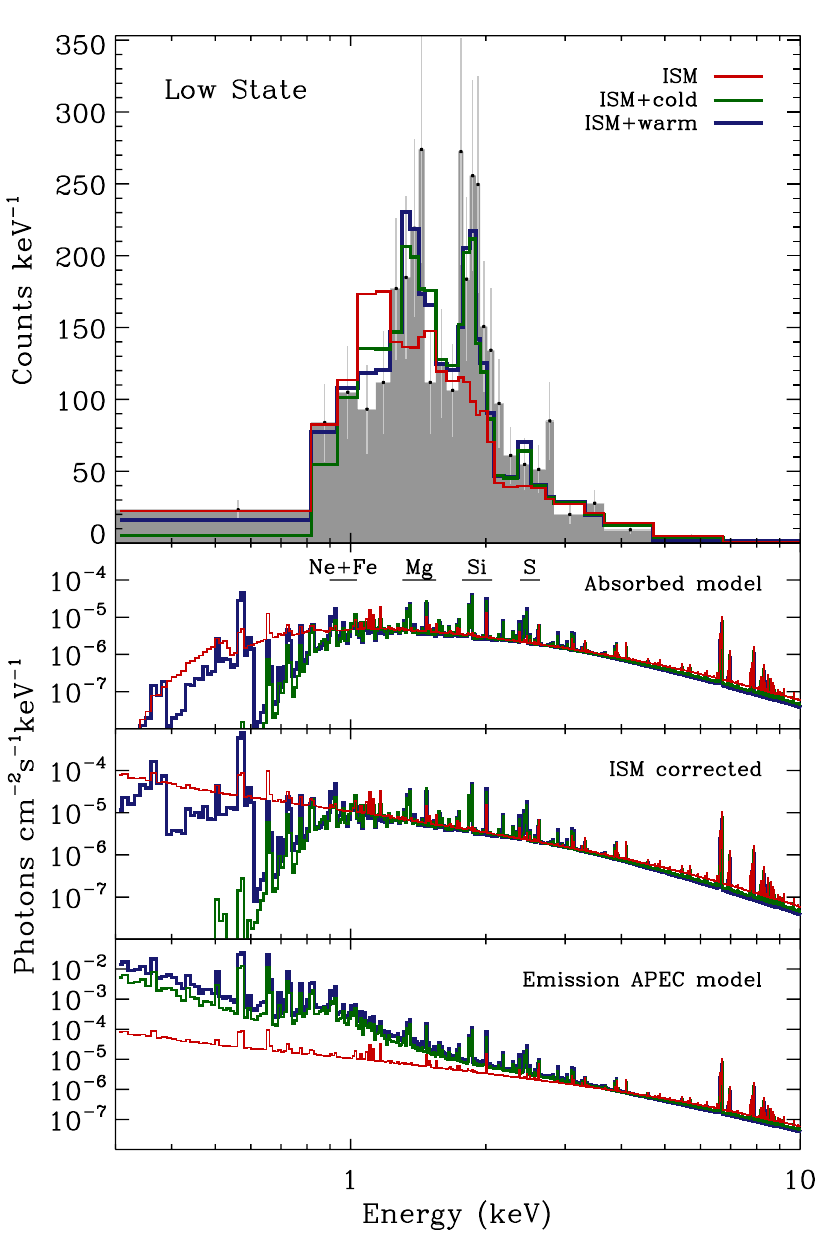}\includegraphics[width=0.5\textwidth]{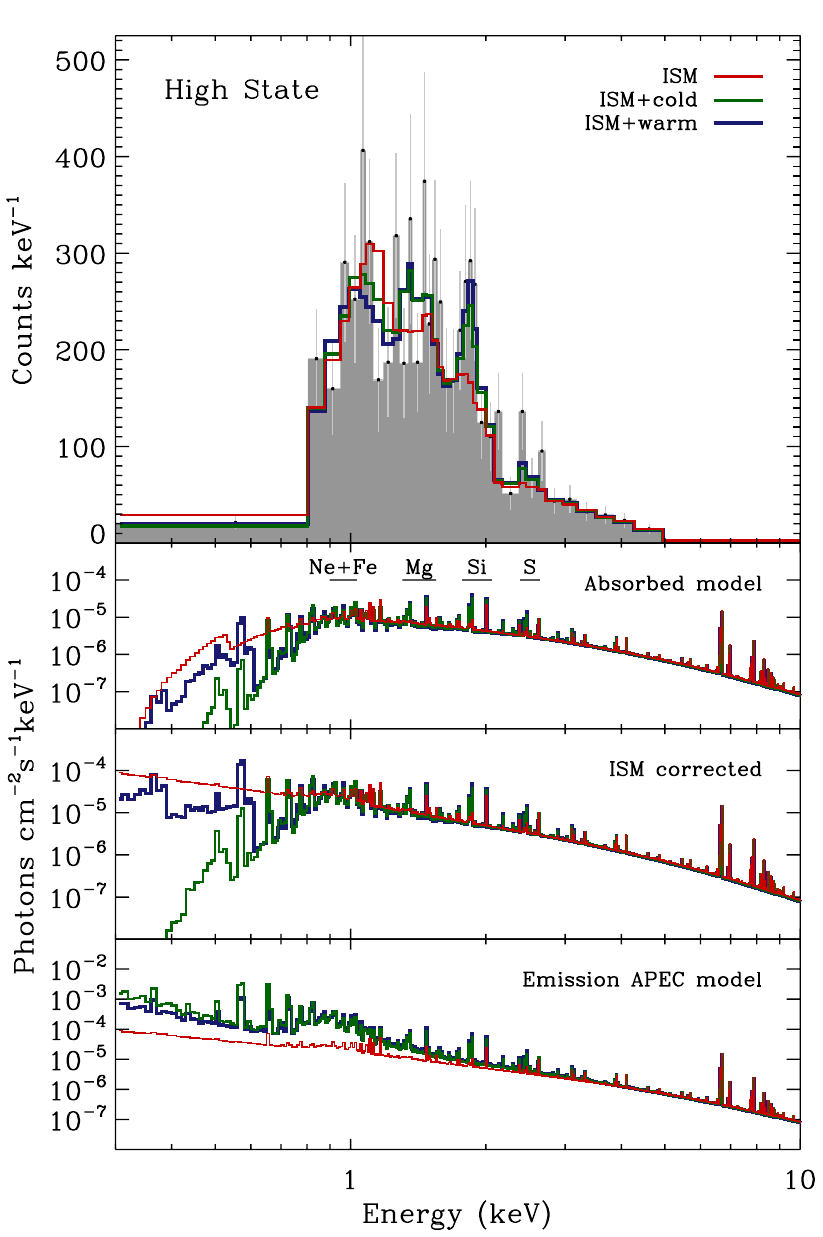}
	\caption{\label{fig:ind}Individual fits to the low state (left) and the high state (right). The emission component of the model is a 4-component fixed-temperature \textit{apec} spectrum (0.2, 0.4, 1.0 and 4.0 keV). The absorption models are a cold ISM only (thin red), a cold ISM plus cold local absorption (green), and a cold ISM plus warm local absorption (thick blue). The three lower panels show, from top to bottom, the best fit for each attenuation model, the best fit local absorption component without the ISM absorption, and the best fit emission component (\textit{apec} only; no ISM/LOC absorption). Note the prominent Mg and Si spectral features between 1 and 2 keV, that can only be reproduced with a relatively cool plasma component.} 
\end{figure*}

We fit the individual models, with the three different absorption models (``ISM only'', ``Cold LOC'', and ``Warm LOC''), to each observation independently. 
The top panels of Fig.~\ref{fig:ind} show the best fit model spectra folded through the ACIS-S response function (red, green and blue lines) overlaid on the observed spectra (grey histogram). 
The 3 bottom panels show, from top to bottom: 
(i) The attenuated model spectra as would be seen from Earth, with dominant spectral features identified.
(ii) The model spectra corrected for the ISM absorption, as emergent from the magnetosphere.
(iii) The emission model spectra (4\textit{T} components; essentially both ISM and LOC corrected) associated with these best fit models. 

The fit results are compiled in Table~\ref{tab:model_ind}. 
Columns 1-3 describe the absorption model under consideration. Columns 4-7 give the normalisation of each \textit{T} component, and column 8 gives the mean plasma temperature weighted by this normalisation. 
The energy fluxes in column 9 are calculated by integrating the model spectra between 0.3 and 10.0 keV. 
We also parametrize the relative hardness of the model spectra by measuring a flux hardness ratio (column 10) that compares soft (0.3-2.0 keV) and hard (2.0-10.0 keV) bands. 
The reduced $\chi^2$ of the model fits are given in column 11.
We also provide the reduced $\chi^2$ values that correspond to the 1$\sigma$, 2$\sigma$, and 3$\sigma$ confidence intervals for the degrees of freedom of each observation (23 and 33 for low state and high state, respectively). 

In the second part of Table~\ref{tab:model_ind}, we compile the characteristics of the the ISM-corrected model spectra (columns 12-15), and those of the emission \textit{4T} model spectra (columns 16-19).
In addition to the flux and hardness ratio described above, we also calculate the associated luminosity in the X-ray band $\log(L_\mathrm{X})$ and the ratio of the X-ray luminosity to the bolometric luminosity $\log(L_\mathrm{X}/L_\mathrm{bol})$.

\begin{table*}
\caption{\label{tab:model_ind}Individual models (shown in Fig.~\ref{fig:ind}). The distance-weighted emission measure normalization of the APEC components ($EM_i$) are in units of $10^{-14}$cm$^{-5}$. These should be multiplied by $4\pi d^2$ to be converted into volume emission measures in cm$^{-3}$. 
We here compile these factors with a greater accuracy than their typical formal modeling uncertainty, in order for our model spectra to be reproducible. The formal error on the Earth-observed flux is or the order of $1\times10^{-14}$\,erg\,s$^{-1}$cm$^{-2}$. All other quantities are given to the significant digit.}
\renewcommand{\arraystretch}{1.2}
\begin{flushleft}
\input{Table3}
\end{flushleft}
\end{table*}

From these models, we reach the following conclusions:
\begin{enumerate}
\item
 \textit{For both observations, the models without local absorption (ISM only) provide a poorer fit than those with  extra local absorption. }
 
In the case of the low state, the improvement in the fit is statistically significant. 
In the case of the high state, all three models are within the 1$\sigma$ confidence interval. 
Indeed, the emission spectra (bottom panels of Fig.~\ref{fig:ind}) and fluxes (column 16 of Table~\ref{tab:model_ind}) of the different models are more similar during the high state observation than during the low state observation. 

The need for this local absorption component can be understood by the fact that a hot plasma component is necessary to explain the hard ($>$3\,keV) tail of the observed spectra. 
However, the model only absorbed by the ISM material is too hot ($kT_\mathrm{avg}$$\simeq$3.7\,keV) to accurately reproduce the very prominent Mg and Si features, which have peak emissivities at cooler temperatures ($\log T$$\simeq$6.8 and 7.0 or $kT$$\simeq$0.5 and 0.8\,keV, respectively). 
The inclusion of a cooler component with a stronger emission measure to model the Mg and Si features cannot reproduce the lack of flux at low energy ($<0.8$\,keV) without the presence of additional absorption.

\item
\textit{During both epochs, the amount of local absorption is very large. }

One explanation for this result could be that the ISM column density has been underestimated. 
However, the $A_\mathrm{V}$ corresponding to these additional column densities would be 11$\pm$3\,mag and 6$\pm$3\,mag for the low and high state, respectively. 
Even during the high state, this difference is larger than the uncertainties on the reddening determined by \mbox{\citetalias{2012MNRAS.425.1278W}} (see Table\,\ref{tab:prop}). 
It is therefore rather unlikely that the extra absorption is of interstellar origin. 

The two models with cold and warm circumstellar absorption result in statistically similar fits. If we consider the likely hypothesis that this absorption is the result of warm, ionised material trapped in the magnetosphere, the column densities needed to reproduce the observed spectra are quite large. As a base for comparison, although the expected wind geometry is different, such a column density would rival the characteristic column density for the non-magnetic early-O supergiants $\zeta$\,Pup or HD\,150136 \citep[see Table\,1 of][]{2010ApJ...719.1767L}. 
However, as pointed out by \citetalias{2012MNRAS.425.1278W}, NGC\,1624-2 has by far the largest H$\alpha$ equivalent width of all the magnetic O-type stars, implying a greater amount of magnetospheric plasma.

\item
\textit{There are large differences in the emission model spectra when comparing the low and high state observations considering the same absorption model.}

The model spectra in the low state are softer than in the high state, as shown in Table \ref{tab:model_ind} by the hardness ratio in column 17 and the $kT_\mathrm{avg}$ in column 8. 
The flux of the emission component is higher during the low state by a factor 3-5. 
The conclusion is the same when considering a cold or a warm circumstellar absorption. The emission spectra corresponding to a different absorption model for a given observation are only slightly different,  as can be seen by comparing the green and blue curves in the bottom panels of Fig.~\ref{fig:ind} (also column 16 and 17 of Table\,\ref{tab:model_ind}), the fluxes and the spectral hardness are different by a factor of $\la2$.

Such a result would be difficult to explain in terms of occultation of a spatially 
stratified temperature distribution. As discussed in the following section,  
simultaneously modelling both epochs with a common emission model provide 
a slightly poorer but still adequate fit to the observations. A high-resolution X-ray 
spectrum of NGC\,1624-2 would provide a useful constraint for the detailed properties
of the magnetosphere emission processes.

\item
\textit{The X-ray emission of NGC\,1624-2 is very luminous, and heavily attenuated.}

Depending on the characteristics of the additional absorption component, the emergent (ISM corrected) X-ray luminosities between 0.3 and 10 keV are $\log(\lx)$ $\sim$32.3 and 32.5 for the low state and high state, corresponding to $\log(\lx/\lbol)$ of \mbox{-6.3} and \mbox{-6.2}, respectively, similar to what is seen in other magnetic O-type stars \citep{2014ApJS..215...10N}.
These models also imply that the shock-heated material produces high intrinsic X-ray fluxes, corresponding to roughly $\log(\lx)$ of $\sim33.7$ and $\sim33.0$, or $\log(\lx/\lbol)$ of \mbox{$-$5.0} and \mbox{$-$5.6}. 
This would imply that 70-95 percent of the MCWS-produced X-ray emission is absorbed before escaping the magnetosphere.  
As a comparison, \citet{2011MNRAS.415.3354C} determined that 80 percent of X-ray emission produced in the wind of the non-magnetic star HD\,93129A, one of the earliest known O stars with a mass-loss of nearly $10^{-5}$M$_\odot$\,yr$^{-1}$, is absorbed before escaping the wind. 
We consider the values derived here to be first order estimates, as the geometry of the models might be too simplified to accurately describe the structure of a real magnetosphere.

\end{enumerate}

\subsection{Results from joint modelling}

Let us now turn our attention to the joint models described previously, where the 4\textit{T} emissivity model is constrained to simultaneously fit both low and high state spectra, allowing only for a change in overall flux reduction during the low state to account for the occultation of the shock-heated plasma by the stellar disk. The results of these joint fits, for the three absorption models, are compiled in Table~\ref{tab:model_joint}. The blank entries represent values that were kept common between the two epochs.

		\begin{table*}
		\caption{\label{tab:model_joint}Same as Table \ref{tab:model_ind} for joint models (Fig.~\ref{fig:joint_comp} shows a comparison between the joint and individual models for the warm local absorption case).}
		\renewcommand{\arraystretch}{1.2}
		\begin{flushleft}
		\input{Table4}
		\end{flushleft}
		\end{table*}

\begin{figure*}
	\includegraphics[width=0.5\textwidth]{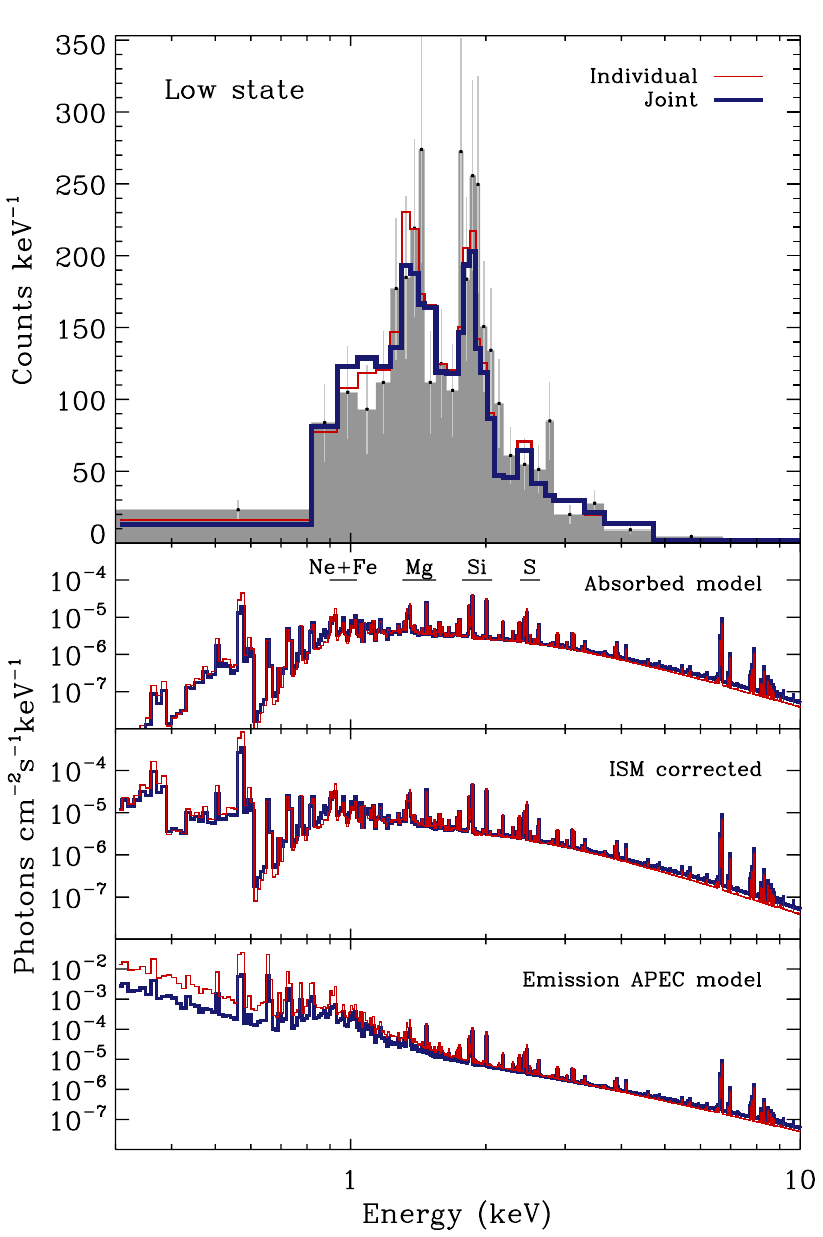}\includegraphics[width=0.5\textwidth]{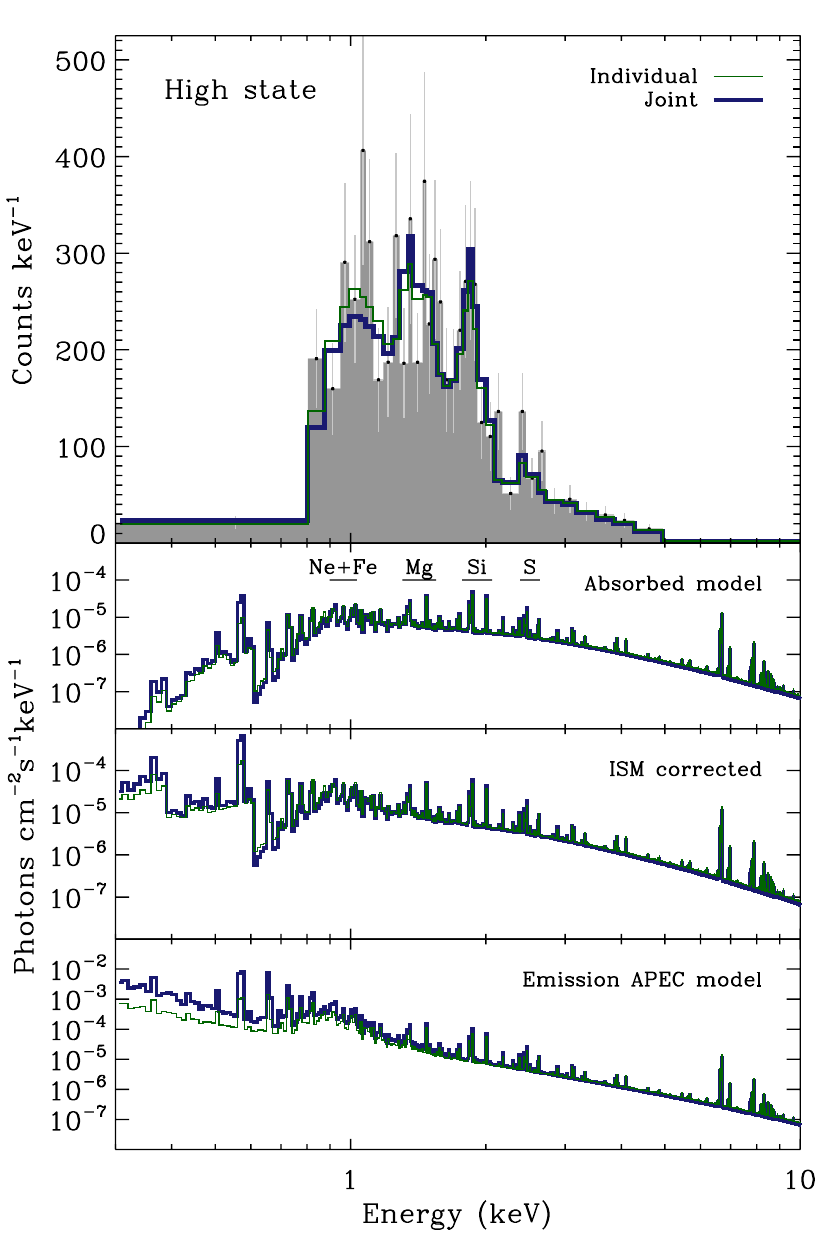}
	\caption{\label{fig:joint_comp} Joint fit of the minimum phase (left) and the maximum phase (right) using a common emission model for both epochs, composed of a 4-temperature \textit{apec} model (with components at 0.2, 0.4, 1.0 and 4.0 keV). Only the overall normalisation of the emission spectrum is allowed to change between the epochs. The absorption model shown here is the "cold ISM with warm local absorption" model. The best-fit joint model is shown in thick blue, compared with the best fits to the observations taken individually from Fig.~\ref{fig:ind} (in thin red for the low state and thin green for the high state) for the same absorption model. The top panel shows the models folded through the instrumental response compared to the data, and the bottom panels shows the models, the models corrected for the ISM absorption, and the emission models (i.e. \textit{4T} only). } 
\end{figure*}

Once again the models with extra local absorption better reproduce the observed spectra, in a statistically significant way for the low state (see column 11).
The joint models provide marginally poorer fits than the individual models, although all the models including local absorption are within the 2$\sigma$ confidence interval.
Fig.~\ref{fig:joint_comp} shows a comparison of the best individual models and the joint model for the warm local absorption model. 
During the low state (left panel), the joint emission model spectrum (bottom, in thick blue) is harder than the best individual emission model (in thin red). During the high state (right panel), the joint emission model spectrum is softer than the best individual emission model (in thin green). This can also be seen by comparing the hardness ratios of the emission models (column 17) and $kT_\mathrm{avg}$ (column 8)  of Table \ref{tab:model_ind} and \ref{tab:model_joint}.

Because of these differences in emission model spectra between the individual and joint approach, the joint model for the low state requires less absorption to reproduce the lack of soft flux (and slightly underestimates the flux in the Mg and Si features) compared to the individual models. 
The opposite effect is seen for the high state. This leads to a less pronounced difference in the column density of the local absorption between the two epochs (column 3).
  
Given the similar quality of the fits from the individual and joint models, the emergent, ISM-corrected fluxes and luminosities are similar (columns 12 and 14). 
Considering now the models with local absorption, the intrinsic flux of the emission component during the high state (column 16) is higher than for the individual models by a factor $\sim$2, leading to an intrinsic luminosity of $\log(\lx)\sim33.4$. With this simplified model, the best value for the flux reduction ($F_\mathrm{min}\sim0.8\,F_\mathrm{max}$) during the minimum phase implies that  \textit{the stellar disk must therefore occult $\sim$20 percent of the X-ray emitting material when the magnetosphere is seen close to edge-on. In addition, the circumstellar material attenuates 85-90\% of the X-ray emission at both phases.}

\section{Discussion}
\label{sec:disc}

\begin{figure}
	\includegraphics[width=0.42\textwidth]{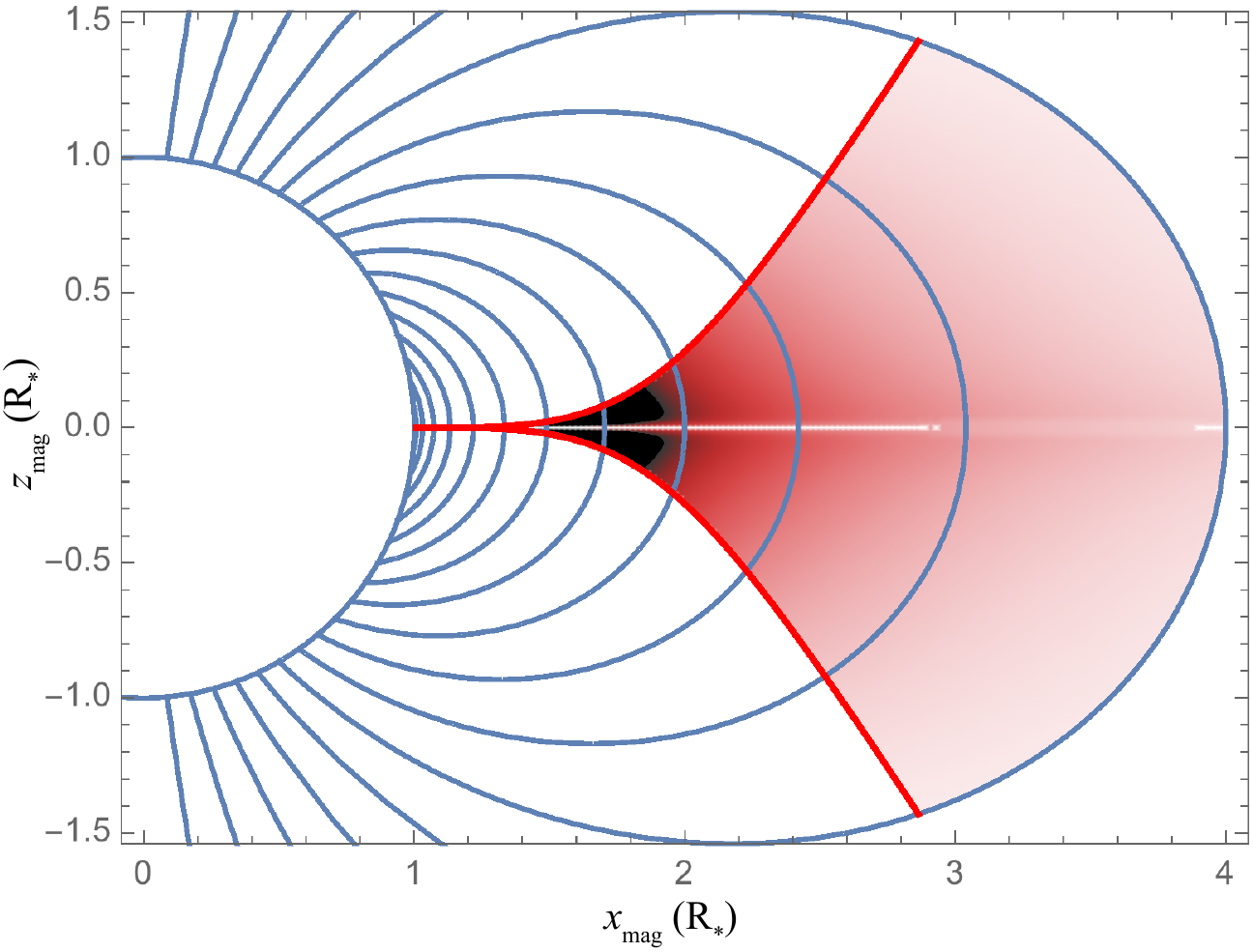}\includegraphics[width=0.06\textwidth]{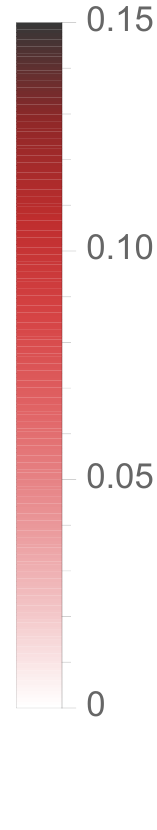}
	\includegraphics[width=0.42\textwidth]{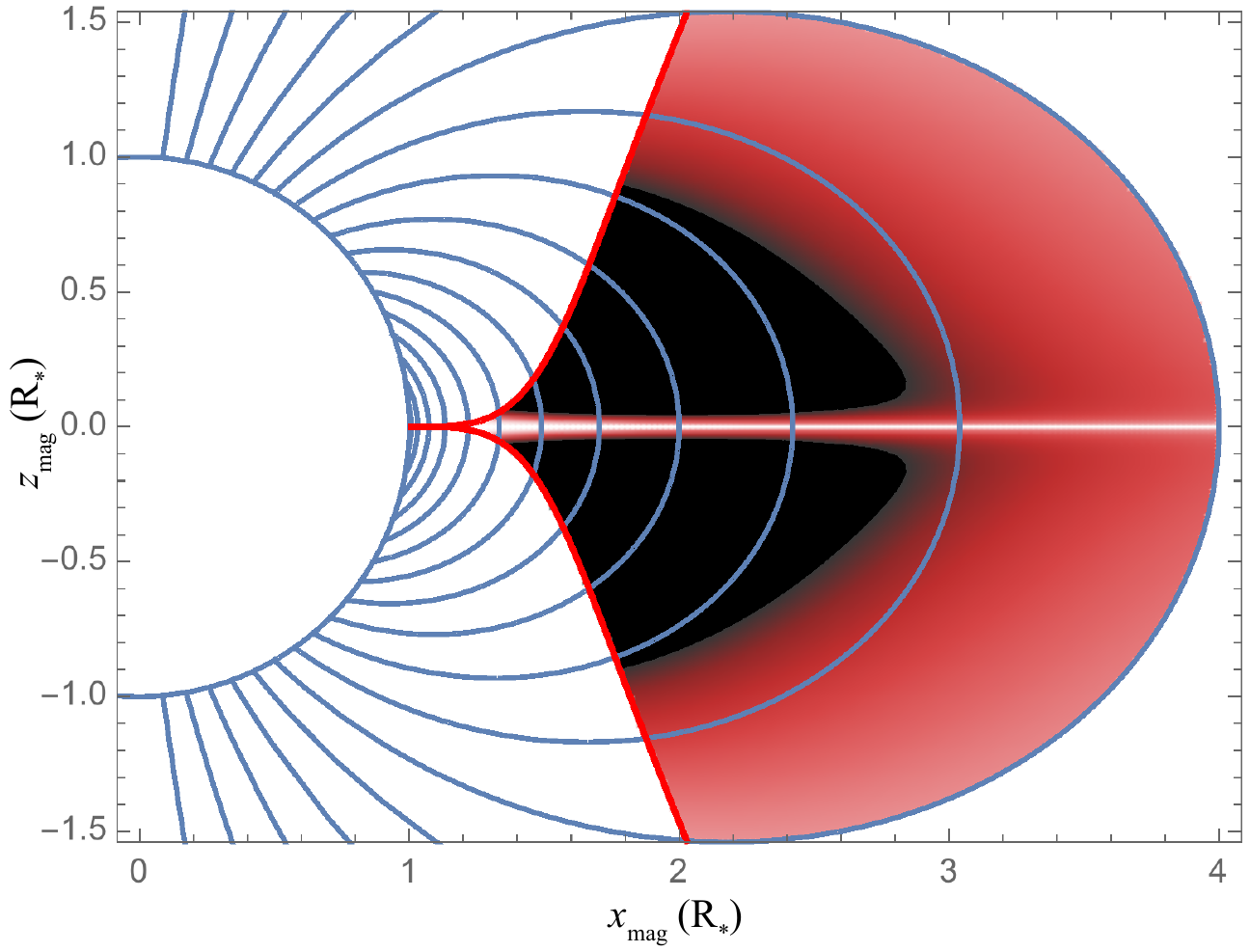}\includegraphics[width=0.06\textwidth]{Fig8bar.pdf}	\caption{\label{fig:adm} Relative spatial distribution of the X-ray emission according to the analytical models of \citet[][see their appendix A]{2014MNRAS.441.3600U} for cooling parameters $\chi_\infty=1$ (top) and $\chi_\infty=10$ (bottom). The shock retreat distributes the X-ray emission over a wider range of latitude. However, the bulk of the X-ray emission is always located relatively close to the star, even for large closure radius (here 4$R_\star$). }
\end{figure}

In the pioneering work of \citet{1997A&A...323..121B}, the X-ray luminosity originating from magnetically confined wind shocks fundamentally scales linearly with wind mass-loss rate, because the thermal energy content of the radiating post-shock plasma depends on the wind kinetic energy flux across the magnetically channeled shock front.
This model was recently refined by \citet{2014MNRAS.441.3600U}, who combined MHD simulations and analytical scalings taking into account the effects of instrument-specific bandpass and ``shock retreat''. 

Shock retreat occurs in low density environments, where the cooling length of the shocked material becomes comparable to the length of the magnetic loop. 
In this case, the X-ray emission is expected to be be spatially extended above and below the magnetic equator, which lowers the pre-shock velocity and thus the kinetic energy available to power the shocks.

The overall importance of the shock-retreat effect can be characterised by the cooling parameter defined in eqn.\ 25 of \citet{2014MNRAS.441.3600U}\footnote{It is important to note that the models of \citet{2014MNRAS.441.3600U} are parametrized by the mass-feeding rate at the base of an \textit{unperturbed} wind, which is not the same as the actual mass-loss ulimately lost by the star.}:
\begin{equation}
\chi_\infty \approx 0.034 \frac{v_8^4 R_{12}}{\dot{M}_{-6}},
\end{equation}
where $v_8\equiv v_\infty/(10^8\,\mathrm{cm}\,\mathrm{s}^{-1})$, $R_{12}\equiv R_\star/(10^{12}\,\mathrm{cm})$, and $\dot{M}_{-6}\equiv \dot{M}/(10^{-6}\,M_\odot\,\mathrm{yr}^{-1})$. 
The case where $\chi>1$ corresponds to a regime where shock retreat is important. The case where $\chi < 1$, represents a regime with dense conditions where the cooling length is small compared to the size of the magnetic loop, 
and the X-ray emission remains from near the loop apex at the magnetic equator.

Using the values listed in Table\,\ref{tab:prop}, we find $\chi_\infty\sim10$ for NGC\,1624-2, implying a moderately strong shock retreat, with associated moderation of X-ray emission level and hardness. Given the strong dependence on the rather uncertain wind terminal speed, a slight reduction to a lower $v_\infty$ value could result in a cooling parameter as low as unity, but likely not much lower. 
Fig.\,\ref{fig:adm} shows the expected spatial distribution of the integrated X-ray emission between 0.3 and 10\,keV from the analytical models described by \citet{2014MNRAS.441.3600U}, for the $\chi_\infty=1$ and $\chi_\infty=10$ cases (see also Owocki et al. in preparation). Although the latitudinal distribution of the X-ray emission is strongly dependent on $\chi_\infty$, its radial extent is concentrated relatively close to the star ($\sim$ 1.5-3 $R_\star$), independent of $\chi_\infty$ as well as the closure radius, located near $R_A$. 
This can be understood as due to the competition between the increase of the velocity shock jump with distance from the stellar surface, and the density decrease due to the areal divergence of the field loop. 

\begin{figure}
	\includegraphics[width=0.5\textwidth]{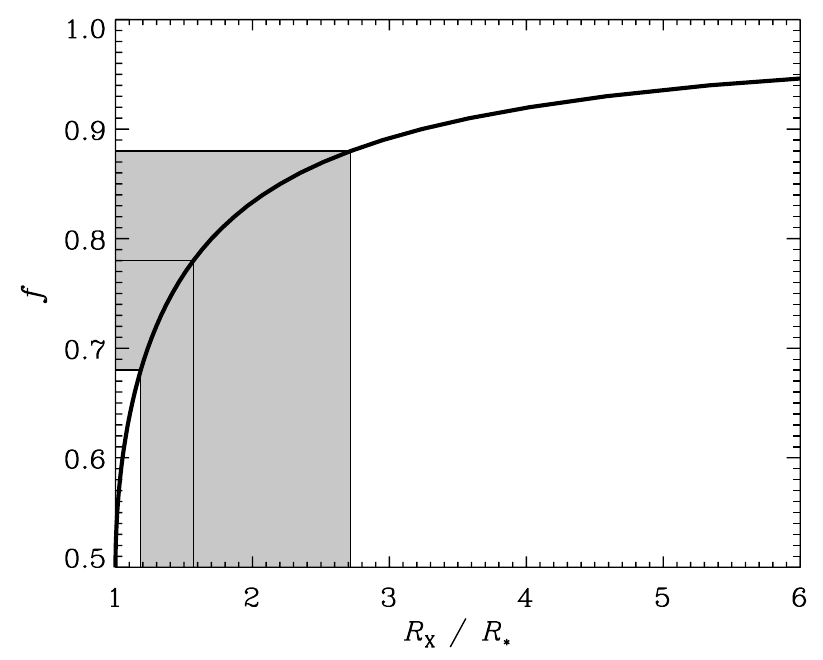}
	\caption{\label{fig:ocult} Reduction of the observed emission ($f=F_\mathrm{obs}/F_\mathrm{tot}$) caused by the occultation by the stellar disk of a ring of X-rays located at a radius $R_\mathrm{X}$ (black thick curve). The thin middle line and shaded area show the constraints obtained by the joint modelling with the ``warm'' circumstellar absorption. This supports the notion that the bulk of the X-rays are emitted relatively close to the stellar surface, as compared to the radius of the outermost closed loops (see text for details).  }
\end{figure}

Our joint modelling of the NGC\,1624-2 spectra suggests that the stellar disk occults $\sim$20 percent of the X-ray emitting material. Using a simple geometric model where the X-ray emission originates from a ring concentrated at a single radius $R_\mathrm{X}$, the maximum occultation (when the magnetosphere is perfectly edge-on) can be expressed as:
\begin{equation}
f \equiv \frac{F_\mathrm{obs}}{F_\mathrm{tot}} = 1- \frac{1}{\pi}\sin^{-1}\left(\frac{R_\star}{R_\mathrm{X}}\right).
\end{equation}
Fig.\,\ref{fig:ocult} shows the constraints on $R_\mathrm{X}$ from this simple occultation model, suggesting that the bulk of the X-ray emission is indeed generated relatively close to the star ($\sim$2$R_\star$), compared to the extent of the largest closed magnetic loops ($\sim$10$R_\star$). 

In the idealized limit of 100\% shock efficiency, the scaling analysis presented by \citet{2014MNRAS.441.3600U} predicts a theoretical X-ray emission for NGC\,1624-2 of $\log(L_\mathrm{X})=34.4 $ in the 0.3 - 10\,keV bandpass. 
By comparison, the empirical analysis presented here gives a factor $\sim10$ lower luminosity for the intrinsic emission component in this bandpass (e.g. $\log L_\mathrm{X}=33.4$ for the joint model with warm local absorption). This indicates a shock efficiency factor of $\sim$10\%, in good agreement with the estimates obtained from comparison with MHD simulations \citep{2014MNRAS.441.3600U}, and also suggested by observations of magnetic massive stars \citep{2014ApJS..215...10N}. 
This supports our conclusion that the X-ray emission from NGC\,1624-2 is highly attenuated, with an amount similar to what is seen in the winds of much earlier non-magnetic O-type stars.

The high intrinsic X-ray luminosity of NGC\,1624-2 is therefore consistent with the strong X-ray emission seen in other magnetic O-type stars. 
Empirically, the observed spectrum of NGC\,1624-2 is hard like that of $\theta^1$\,Ori\,C; but intrinsically it is softer, making it more similar to the other Of?p stars. 
On the other hand, NGC\,1624-2 is distinguished from the other Of?p stars by its large (phase-dependent) absorption, due to its spatially large magnetosphere. 

Direct MHD simulations for high degrees of magnetic wind confinement, as are needed for NGC\,1624-2, are not computationally feasible at this time, although new developments in rigid-field hydrodynamics models might be available in a near future \citep{2007MNRAS.382..139T,2015IAUS..307..449B}. 
However, we can use the 3-D MHD simulation presented by \citet{2013MNRAS.428.2723U} for a lower magnetic wind confinement, similar to that found for other magnetic O-type stars, as a baseline to explore the level and modulation amplitude of X-ray absorbing column densities.

\begin{figure*}
	\includegraphics[width=0.5\textwidth]{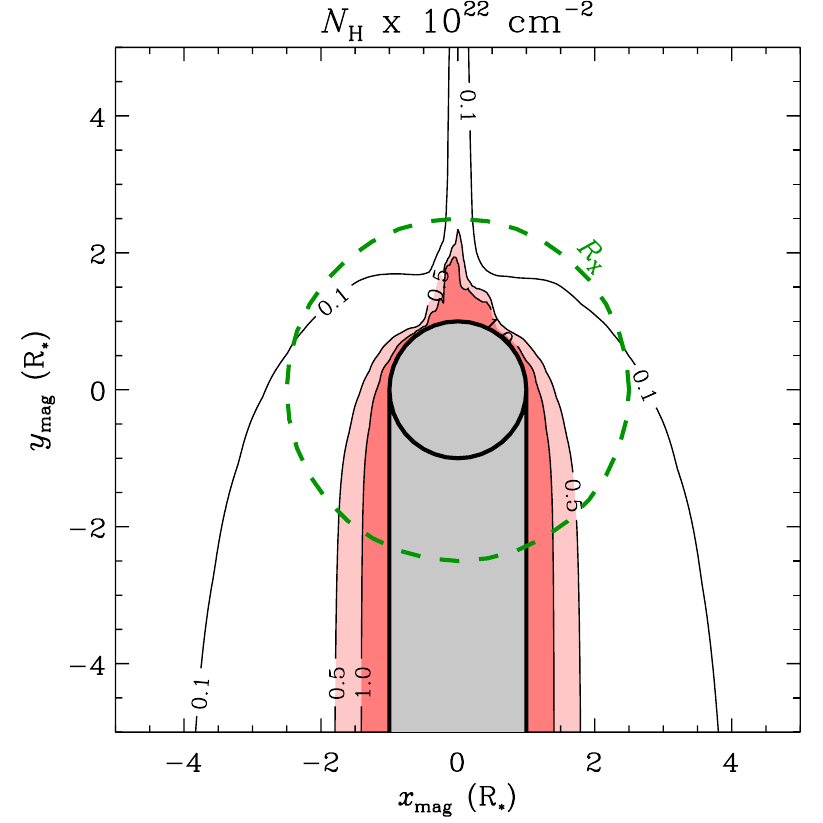}\includegraphics[width=0.5\textwidth]{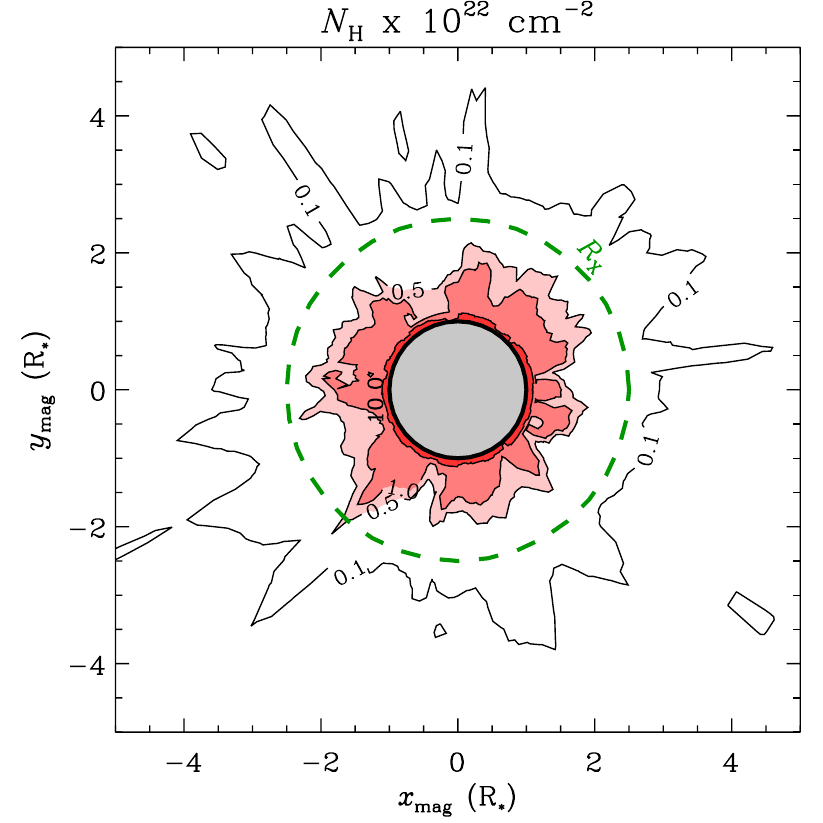}
	\caption{\label{fig:mhd} Time-averaged (50\,ks) column density $n_H$ between any location \textit{on the magnetic equatorial plane} and the observer, calculated from the 3-D MHD simulation presented by \citet{2013MNRAS.428.2723U} for a low magnetic wind confinement, similar to that found for other magnetic O-type stars. In both panel, the magnetic equatorial plane is in the plane of the page, and the magnetic pole is out of the page. The right panel correspond to the high state when the magnetosphere is viewed pole-on (the observer is  located in front of the page), and the left panel correspond to the low state when the magnetosphere is viewed edge-on (observer is located at the top of the page, looking into the page). The grey circle is the stellar disk and the grey zone at the bottom of the right panel marks regions occulted by the stellar disk. The dashed green circle is located at $2R_\star$.}
\end{figure*}

Fig.\,\ref{fig:mhd} shows contour plots of the column density $N_H$, time-averaged over 50\,ks, between any location \textit{in the magnetic equatorial plane} and the observer. 
The right panel corresponds to the high state, when the magnetosphere is viewed pole-on. The observer is therefore located in front of the page. 
The left panel corresponds to the low state, when the magnetosphere is viewed edge-on -- the observer is therefore located at the top of the page and viewing in the plane of the page toward the bottom. The grey region at the bottom marks the occultation of a portion of the equatorial plane by the stellar disk.
The green dashed circles mark $2~R_\star$, the expected location of the bulk of the X-ray emission.

This 3D simulation predicts a column density between 0.1-0.5$\times10^{22}$\,cm$^{-2}$ between the observer and X-ray emission originating at $\sim2~R_\star$. This is consistent with the low absorption column observed in other magnetic O-type stars \citep[e.g.][]{2005ApJ...628..986G,2014ApJS..215...10N}. Regions of higher column density exist along the green circle for the low state on the far side of the magnetic equatorial plane, but these regions are already occulted by the stellar disk. It is therefore likely that no large changes in absorption will be observed for stars with magnetospheres significantly smaller than that of NGC\,1624-2.

The empirical column density for NGC\,1624-2 is higher (1-4$\times10^{22}$\,cm$^{-2}$), which, in these simulations, only occurs for rays that impact very close to the star. A suitably larger magnetosphere could produce such a larger attenuation, as a larger fraction of the wind will be channeled into closed magnetic loops, increasing the overall column density. 
An extension of the X-ray scaling analysis of \citet{2014MNRAS.441.3600U} to account for the spatial distribution of pre-shock and cool post-shock material could help quantify the X-ray absorption for large magnetospheres, currently out of reach of present MHD simulations (Owocki et al. in prep).

\section{Conclusion}

We observed NGC\,1624-2, the O-type star with the largest known magnetic field ($B_p\sim{20}$\,kG) using the ACIS-S camera on-board the \textit{Chandra} X-ray Observatory. 
Our two observations were obtained at the low and high state of the periodic H$\alpha$ emission cycle, corresponding to the rotational phases where the magnetic field is the closest to equator-on and pole-on, respectively.

In both observations most of the photon counts have energies higher than 0.8 keV.
No evidence of short term variability is found within each exposure. However, the net photon flux is 40 percent higher during the H$\alpha$ emission high state, which shows an excess of softer ($\sim$1\,keV) photons compared to the low state.

We model the spectra with optically thin, collisional radiative plasma models from the Astrophysical Plasma Emission Code. 
We find that for both observations a large local magnetospheric absorption, in addition to that from the interstellar medium, provides a much better fit to the features present in the spectra.

The X-ray emission of NGC\,1624-2 is therefore very luminous, and heavily attenuated, by $\sim 70-95$ percent. 
Therefore, although the spectrum of NGC\,1624-2 appears hard like that of the archetypical magnetic star $\theta^1$\,Ori\,C, it is in fact intrinsically softer and more similar to the other magnetic Of?p stars. 
However, the large amount of circumstellar absorption present in NGC\,1624-2 is not seen in any of the other magnetic stars. We argue that such a high attenuation, and its observed variation with stellar rotation, is consistent with a larger magnetic closure radius, within the framework of a dynamical magnetosphere in the Magnetically Confined Wind Shock X-ray context, and consistent with its remarkably strong magnetic field.

The case of NGC\,1624-2 also revealed that the plasma postshock temperature, and therefore the intrinsic X-ray spectral hardness, should not depend on the confinement level. The question thus remains as to why the emission of the archetype magnetic O-type star $\theta^1$\,Ori\,C is harder than that of the other magnetic O-type stars observed in X-rays.
Upcoming HST/COS observations will address the ionisation structure and kinematics of the plasma, as well as the feeding and mass-loss rates of the giant magnetosphere of NGC\,1624-2. These new observations may reveal important information about the pre-shock wind velocity, that can be compared to $\theta^1$\,Ori\,C to verify whether the velocity structure of the magnetosphere is responsible for this difference is plasma temperature.

\section*{Acknowledgments}

Support for this work was provided by NASA through Chandra Award G03-14017C issued to the Florida Institute of Technology by the Chandra X-ray Observatory Center which is operated by the Smithsonian Astrophysical Observatory for and behalf of NASA under contract NAS8- 03060.
DHC acknowledges support from SAO Chandra grants TM4-15001B, G03-14017B, and AR2-13001A.
GAW acknowledges Discovery Grant support from the Natural Science and Engineering Research Council of Canada (NSERC).
YN acknowledges support from  the Fonds National de la Recherche Scientifique (Belgium), the PRODEX XMM contract and and ARC grant for concerted research action financed by the federation Wallonia-Brussels.
SPO acknowledges NASA ATP grant NNX11AC40G and NNX12AC72G.
JOS acknowledges support from SAO Chandra grant TM3-14001A. 
AuD acknowledges support from SAO Chandra grant TM4-15001A, and NASA ATP grant NNX12AC72G.
AuD also acknowledges support for Program number HST-GO-13629.008-A provided by NASA through a grant from the Space Telescope Science Institute, which is operated by the Association of Universities for Research in Astronomy, Incorporated, under NASA contract NAS5-26555.
MRG acknowledges support from SAO Chandra grants AR3-14001A.
The authors thank the referee for useful comments and suggestions, and acknowledge help from and discussion with N. Walborn, R. Townsend, R. H. Barb\'a, G. Rauw, J. Ma{\'{\i}}z Apell{\'a}niz, and A. Pellerin.

\bibliographystyle{mn2e_fix2}
\bibliography{database}

\label{lastpage}
\end{document}

%% file: aas_macros.tex
%
%
%


\def\jnl@style{\it}
\def\aaref@jnl#1{{\jnl@style#1}}

\def\aaref@jnl#1{{\jnl@style#1}}

\def\aj{\aaref@jnl{AJ}}                   
\def\araa{\aaref@jnl{ARA\&A}}             
\def\apj{\aaref@jnl{ApJ}}                 
\def\apjl{\aaref@jnl{ApJ}}                
\def\apjs{\aaref@jnl{ApJS}}               
\def\ao{\aaref@jnl{Appl.~Opt.}}           
\def\apss{\aaref@jnl{Ap\&SS}}             
\def\aap{\aaref@jnl{A\&A}}                
\def\aapr{\aaref@jnl{A\&A~Rev.}}          
\def\aaps{\aaref@jnl{A\&AS}}              
\def\azh{\aaref@jnl{AZh}}                 
\def\baas{\aaref@jnl{BAAS}}               
\def\jrasc{\aaref@jnl{JRASC}}             
\def\memras{\aaref@jnl{MmRAS}}            
\def\mnras{\aaref@jnl{MNRAS}}             
\def\pra{\aaref@jnl{Phys.~Rev.~A}}        
\def\prb{\aaref@jnl{Phys.~Rev.~B}}        
\def\prc{\aaref@jnl{Phys.~Rev.~C}}        
\def\prd{\aaref@jnl{Phys.~Rev.~D}}        
\def\pre{\aaref@jnl{Phys.~Rev.~E}}        
\def\prl{\aaref@jnl{Phys.~Rev.~Lett.}}    
\def\pasp{\aaref@jnl{PASP}}               
\def\pasj{\aaref@jnl{PASJ}}               
\def\qjras{\aaref@jnl{QJRAS}}             
\def\skytel{\aaref@jnl{S\&T}}             
\def\solphys{\aaref@jnl{Sol.~Phys.}}      
\def\sovast{\aaref@jnl{Soviet~Ast.}}      
\def\ssr{\aaref@jnl{Space~Sci.~Rev.}}     
\def\zap{\aaref@jnl{ZAp}}                 
\def\nat{\aaref@jnl{Nature}}              
\def\iaucirc{\aaref@jnl{IAU~Circ.}}       
\def\aplett{\aaref@jnl{Astrophys.~Lett.}} 
\def\apspr{\aaref@jnl{Astrophys.~Space~Phys.~Res.}}
\def\bain{\aaref@jnl{Bull.~Astron.~Inst.~Netherlands}} 
\def\fcp{\aaref@jnl{Fund.~Cosmic~Phys.}}  
\def\gca{\aaref@jnl{Geochim.~Cosmochim.~Acta}}   
\def\grl{\aaref@jnl{Geophys.~Res.~Lett.}} 
\def\jcp{\aaref@jnl{J.~Chem.~Phys.}}      
\def\jgr{\aaref@jnl{J.~Geophys.~Res.}}    
\def\jqsrt{\aaref@jnl{J.~Quant.~Spec.~Radiat.~Transf.}}
\def\memsai{\aaref@jnl{Mem.~Soc.~Astron.~Italiana}}
\def\nphysa{\aaref@jnl{Nucl.~Phys.~A}}   
\def\physrep{\aaref@jnl{Phys.~Rep.}}   
\def\physscr{\aaref@jnl{Phys.~Scr}}   
\def\planss{\aaref@jnl{Planet.~Space~Sci.}}   
\def\procspie{\aaref@jnl{Proc.~SPIE}}   

\let\astap=\aap
\let\apjlett=\apjl
\let\apjsupp=\apjs
\let\applopt=\ao

%% file: Table2.tex
\begin{tabular}{ l  c  c  }
\hline
\mcc{} & \mcc{H$\alpha$ low state} & \mcc{H$\alpha$ high state} \\
\hline
ObsID & 14572 & 14571 \\ 
Date & 2013-08-03 & 	2013-10-27 \\ 
Start HJD & 2456508.20 & 2456592.84 \\ 
Rotational phase & 0.43$\pm$0.09 & 0.96$\pm$0.09 \\ 
Net exposure (ks) & 49.5 & 49.5 \\ 
Net counts &301$^{+29}_{-29}$ &431$^{+35}_{-35}$ \\
Count rate (ct\,s$^{-1}$) &6.1$^{+0.6}_{-0.6} \times10^{-3}$ &8.7$^{+0.7}_{-0.7} \times10^{-3}$ \\
Photon flux (Ph\,cm$^{-2}$s$^{-1}$) &1.7$^{+0.2}_{-0.2} \times10^{-5}$ &2.4$^{+0.2}_{-0.2} \times10^{-5}$ \\
\hline
\end{tabular}

%% file: Table3.tex
\begin{tabular}{ l  c  c  c  c  c  c  c  c  c  c  }
\hline
\mcc{Model} & \mcc{$nH^{\mathrm{ISM}}$} & \mcc{$nH^{\mathrm{LOC}}$} & \mcc{$EM_1$} & \mcc{$EM_2$} & \mcc{$EM_3$} & \mcc{$EM_4$} & \mcc{$kT_\mathrm{avg}$} & \mcc{Flux} & \mcc{$H/S$} & \mcc{$\chi^2_\mathrm{red}$} \\
\mcc{} & \mcc{10$^{22}$cm$^{-2}$} & \mcc{10$^{22}$cm$^{-2}$} & \mcc{(0.2\,keV)} & \mcc{(0.6\,keV)} & \mcc{(1.0\,keV)} & \mcc{(4.0\,keV)} & \mcc{(keV)} & \mcc{erg\,s$^{-1}$cm$^{-2}$} & \mcc{} & \mcc{} \\
\mcc{(1)} & \mcc{(2)} & \mcc{(3)} & \mcc{(4)} & \mcc{(5)} & \mcc{(6)} & \mcc{(7)} & \mcc{(8)} & \mcc{(9)} & \mcc{(10)} & \mcc{(11)} \\
\hline
\multicolumn{11}{c}{Low state~~~$\chi^2_\mathrm{red}$ (1$\sigma$,2$\sigma$,3$\sigma$)=1.12, 1.55, 2.00 for 23 DF} \\
ISM only & 0.48 & 0 & 3.59E-06 & 0 & 0 & 4.06E-05 & 3.7 & 4.7E-14 & 2.5 & 2.12 \\ 
Cold LOC & 0.48 & 1.78$\pm$0.50 & 6.83E-04 & 4.28E-05 & 6.14E-05 & 3.18E-05 & 0.4 & 4.6E-14 & 2.3 & 1.52 \\ 
Warm LOC & 0.48 & 3.93$\pm$0.73 & 1.91E-03 & 9.85E-09 & 1.12E-04 & 2.71E-05 & 0.3 & 4.6E-14 & 2.0 & 1.14 \\ 
\hline
\multicolumn{11}{c}{High state~~~$\chi^2_\mathrm{red}$ (1$\sigma$,2$\sigma$,3$\sigma$)=1.10, 1.45, 1.81 for 33 DF} \\
ISM only & 0.48 & 0 & 0 & 0 & 4.14E-06 & 5.82E-05 & 3.8 & 7.0E-14 & 2.3 & 1.09 \\ 
Cold LOC & 0.48 & 0.92$\pm$0.44 & 1.78E-04 & 2.33E-05 & 3.65E-05 & 5.61E-05 & 1.1 & 7.1E-14 & 2.3 & 0.82 \\ 
Warm LOC & 0.48 & 1.96$\pm$0.46 & 5.71E-05 & 3.34E-05 & 4.98E-05 & 5.46E-05 & 1.5 & 7.1E-14 & 2.2 & 0.79 \\ 
\hline
\end{tabular}
\begin{tabular}{ l  c  c  c  c  c  c  c  c  }
\hline
\mcc{} & \multicolumn{4}{c}{ISM corrected} & \multicolumn{4}{c}{Emission component only} \\
\mcc{Model} & \mcc{Flux} & \mcc{$H/S$} & \mcc{$\log(L_\mathrm{X})$} & \mcc{$\log(L_\mathrm{X}/L_\mathrm{bol})$} & \mcc{Flux} & \mcc{$H/S$} & \mcc{$\log(L_\mathrm{X})$} & \mcc{$\log(L_\mathrm{X}/L_\mathrm{bol})$} \\
\mcc{} & \mcc{erg\,s$^{-1}$cm$^{-2}$} & \mcc{} & \mcc{erg\,s$^{-1}$} & \mcc{} & \mcc{erg\,s$^{-1}$cm$^{-2}$} & \mcc{} & \mcc{erg\,s$^{-1}$} & \mcc{} \\
\mcc{(1)} & \mcc{(12)} & \mcc{(13)} & \mcc{(14)} & \mcc{(15)} & \mcc{(16)} & \mcc{(17)} & \mcc{(18)} & \mcc{(19)} \\
\hline
\multicolumn{9}{c}{Low state} \\
ISM only & 7.0E-14 & 1.0 & 32.3 & -6.3 & 7.0E-14 & 1.0 & 32.3 & -6.3 \\ 
Cold LOC & 5.5E-14 & 1.5 & 32.2 & -6.4 & 1.1E-12 & 0.04 & 33.6 & -5.1 \\ 
Warm LOC & 7.3E-14 & 0.8 & 32.4 & -6.3 & 2.6E-12 & 0.02 & 33.9 & -4.8 \\ 
\hline
\multicolumn{9}{c}{High state} \\
ISM only & 1.0E-13 & 1.0 & 32.5 & -6.2 & 1.0E-13 & 1.0 & 32.5 & -6.2 \\ 
Cold LOC & 9.0E-14 & 1.3 & 32.5 & -6.2 & 4.4E-13 & 0.1 & 33.1 & -5.5 \\ 
Warm LOC & 9.6E-14 & 1.1 & 32.5 & -6.2 & 3.4E-13 & 0.2 & 33.0 & -5.7 \\ 
\hline
\end{tabular}

%% file: Table4.tex
\begin{tabular}{ l  c  c  c  c  c  c  c  c  c  c  }
\hline
\mcc{Model} & \mcc{$nH^{\mathrm{ISM}}$} & \mcc{$nH^{\mathrm{LOC}}$} & \mcc{$EM_1$} & \mcc{$EM_2$} & \mcc{$EM_3$} & \mcc{$EM_4$} & \mcc{$kT_\mathrm{avg}$} & \mcc{Flux} & \mcc{$H/S$} & \mcc{$\chi^2_\mathrm{red}$} \\
\mcc{} & \mcc{10$^{22}$cm$^{-2}$} & \mcc{10$^{22}$cm$^{-2}$} & \mcc{(0.2\,keV)} & \mcc{(0.6\,keV)} & \mcc{(1.0\,keV)} & \mcc{(4.0\,keV)} & \mcc{(keV)} & \mcc{erg\,s$^{-1}$cm$^{-2}$} & \mcc{} & \mcc{} \\
\mcc{(1)} & \mcc{(2)} & \mcc{(3)} & \mcc{(4)} & \mcc{(5)} & \mcc{(6)} & \mcc{(7)} & \mcc{(8)} & \mcc{(9)} & \mcc{(10)} & \mcc{(11)} \\
\hline
\multicolumn{11}{c}{Low state~~~$\chi^2_\mathrm{red}$ (1$\sigma$,2$\sigma$,3$\sigma$)=1.12, 1.55, 2.00 for 23 DF} \\
ISM only &  & 0 &  & & & &  & 4.6E-14 & 2.5 & 2.28 \\ 
Cold LOC &  & 1.41$\pm$0.36 &  & & & &  & 4.9E-14 & 2.6 & 1.68 \\ 
Warm LOC &  & 3.02$\pm$0.40 &  & & & &  & 4.9E-14 & 2.4 & 1.35 \\ 
\hline
\multicolumn{11}{c}{High state~~~$\chi^2_\mathrm{red}$ (1$\sigma$,2$\sigma$,3$\sigma$)=1.10, 1.45, 1.81 for 33 DF} \\
ISM only & 0.48 & 0 & 0 & 0 & 1.57E-06 & 6.25E-05 & 3.9 & 7.3E-14 & 2.496 & 1.13 \\ 
Cold LOC & 0.48 & 1.13$\pm$0.36 & 3.57E-04 & 3.83E-05 & 4.63E-05 & 5.06E-05 & 0.7 & 6.8E-14 & 2.077 & 0.85 \\ 
Warm LOC & 0.48 & 2.61$\pm$0.40 & 4.39E-04 & 3.61E-05 & 8.29E-05 & 4.61E-05 & 0.6 & 6.8E-14 & 1.994 & 0.86 \\ 
\hline
\end{tabular}
\begin{tabular}{ l  c  c  c  c  c  c  c  c  }
\hline
\mcc{} & \multicolumn{4}{c}{ISM corrected} & \multicolumn{4}{c}{Emission component only} \\
\mcc{Model} & \mcc{Flux} & \mcc{$H/S$} & \mcc{$\log(L_\mathrm{X})$} & \mcc{$\log(L_\mathrm{X}/L_\mathrm{bol})$} & \mcc{Flux} & \mcc{$H/S$} & \mcc{$\log(L_\mathrm{X})$} & \mcc{$\log(L_\mathrm{X}/L_\mathrm{bol})$} \\
\mcc{} & \mcc{erg\,s$^{-1}$cm$^{-2}$} & \mcc{} & \mcc{erg\,s$^{-1}$} & \mcc{} & \mcc{erg\,s$^{-1}$cm$^{-2}$} & \mcc{} & \mcc{erg\,s$^{-1}$} & \mcc{} \\
\mcc{(1)} & \mcc{(12)} & \mcc{(13)} & \mcc{(14)} & \mcc{(15)} & \mcc{(16)} & \mcc{(17)} & \mcc{(18)} & \mcc{(19)} \\
\hline
\multicolumn{9}{c}{Low state} \\
ISM only & 6.5E-14 & 1.1 & 32.3 & -6.4 & 0.6 $F_\mathrm{high}$ &  & 32.3 & -6.4 \\ 
Cold LOC & 5.9E-14 & 1.7 & 32.3 & -6.4 & 0.8 $F_\mathrm{high}$ &  & 33.2 & -5.4 \\ 
Warm LOC & 6.7E-14 & 1.1 & 32.3 & -6.4 & 0.8 $F_\mathrm{high}$ &  & 33.3 & -5.3 \\ 
\hline
\multicolumn{9}{c}{High state} \\
ISM only & 1.0E-13 & 1.1 & 32.5 & -6.2 & 1.0E-13 & 1.1 & 32.5 & -6.2 \\ 
Cold LOC & 8.6E-14 & 1.3 & 32.4 & -6.2 & 7.1E-13 & 0.08 & 33.4 & -5.3 \\ 
Warm LOC & 1.0E-13 & 0.9 & 32.5 & -6.2 & 8.7E-13 & 0.07 & 33.4 & -5.2 \\ 
\hline
\end{tabular}

%% file: Dash2.bbl
\begin{thebibliography}{}

\bibitem[\protect\citeauthoryear{{Antokhin}, {Owocki} \& {Brown}}{{Antokhin}
  et~al.}{2004}]{2004ApJ...611..434A}
{Antokhin} I.~I.,  {Owocki} S.~P.,    {Brown} J.~C.,  2004, \apj, 611, 434

\bibitem[\protect\citeauthoryear{{Arnaud}}{{Arnaud}}{1996}]{1996ASPC..101...17A}
{Arnaud} K.~A.,  1996, in {G.~H.~Jacoby \& J.~Barnes} ed.,  Astronomical
  Society of the Pacific Conference Series Vol. 101, Astronomical Data Analysis
  Software and Systems V. p.~17

\bibitem[\protect\citeauthoryear{{Asplund}, {Grevesse}, {Sauval} \&
  {Scott}}{{Asplund} et~al.}{2009}]{2009ARA&A..47..481A}
{Asplund} M.,  {Grevesse} N.,  {Sauval} A.~J.,    {Scott} P.,  2009, \araa, 47,
  481

\bibitem[\protect\citeauthoryear{{Babel} \& {Montmerle}}{{Babel} \&
  {Montmerle}}{1997a}]{1997ApJ...485L..29B}
{Babel} J.,  {Montmerle} T.,  1997a, \apjl, 485, L29

\bibitem[\protect\citeauthoryear{{Babel} \& {Montmerle}}{{Babel} \&
  {Montmerle}}{1997b}]{1997A&A...323..121B}
{Babel} J.,  {Montmerle} T.,  1997b, \aap, 323, 121

\bibitem[\protect\citeauthoryear{{Bard} \& {Townsend}}{{Bard} \&
  {Townsend}}{2015}]{2015IAUS..307..449B}
{Bard} C.,  {Townsend} R.,  2015, in IAU Symposium. pp 449--450

\bibitem[\protect\citeauthoryear{{Cantiello} et~al.,}{{Cantiello}
  et~al.}{2009}]{2009A&A...499..279C}
{Cantiello} M.  et~al., 2009, \aap, 499, 279

\bibitem[\protect\citeauthoryear{{Cohen}, {Gagn{\'e}}, {Leutenegger},
  {MacArthur}, {Wollman}, {Sundqvist}, {Fullerton} \& {Owocki}}{{Cohen}
  et~al.}{2011}]{2011MNRAS.415.3354C}
{Cohen} D.~H.,  {Gagn{\'e}} M.,  {Leutenegger} M.~A.,  {MacArthur} J.~P.,
  {Wollman} E.~E.,  {Sundqvist} J.~O.,  {Fullerton} A.~W.,    {Owocki} S.~P.,
  2011, \mnras, 415, 3354

\bibitem[\protect\citeauthoryear{{Donati}, {Babel}, {Harries}, {Howarth},
  {Petit} \& {Semel}}{{Donati} et~al.}{2002}]{2002MNRAS.333...55D}
{Donati} J.-F.,  {Babel} J.,  {Harries} T.~J.,  {Howarth} I.~D.,  {Petit} P.,
   {Semel} M.,  2002, \mnras, 333, 55

\bibitem[\protect\citeauthoryear{{Donati}, {Howarth}, {Bouret}, {Petit},
  {Catala} \& {Landstreet}}{{Donati} et~al.}{2006}]{2006MNRAS.365L...6D}
{Donati} J.-F.,  {Howarth} I.~D.,  {Bouret} J.-C.,  {Petit} P.,  {Catala} C.,
   {Landstreet} J.,  2006, \mnras, 365, L6

\bibitem[\protect\citeauthoryear{{Foster}, {Ji}, {Smith} \&
  {Brickhouse}}{{Foster} et~al.}{2012}]{2012ApJ...756..128F}
{Foster} A.~R.,  {Ji} L.,  {Smith} R.~K.,    {Brickhouse} N.~S.,  2012, \apj,
  756, 128

\bibitem[\protect\citeauthoryear{{Gagn{\'e}}, {Caillault}, {Stauffer} \&
  {Linsky}}{{Gagn{\'e}} et~al.}{1997}]{1997ApJ...478L..87G}
{Gagn{\'e}} M.,  {Caillault} J.-P.,  {Stauffer} J.~R.,    {Linsky} J.~L.,
  1997, \apjl, 478, L87

\bibitem[\protect\citeauthoryear{{Gagn{\'e}}, {Oksala}, {Cohen}, {Tonnesen},
  {ud-Doula}, {Owocki}, {Townsend} \& {MacFarlane}}{{Gagn{\'e}}
  et~al.}{2005}]{2005ApJ...628..986G}
{Gagn{\'e}} M.,  {Oksala} M.~E.,  {Cohen} D.~H.,  {Tonnesen} S.~K.,  {ud-Doula}
  A.,  {Owocki} S.~P.,  {Townsend} R.~H.~D.,    {MacFarlane} J.~J.,  2005,
  \apj, 628, 986

\bibitem[\protect\citeauthoryear{{G{\"u}ver} \& {{\"O}zel}}{{G{\"u}ver} \&
  {{\"O}zel}}{2009}]{2009MNRAS.400.2050G}
{G{\"u}ver} T.,  {{\"O}zel} F.,  2009, \mnras, 400, 2050

\bibitem[\protect\citeauthoryear{{Howarth} et~al.,}{{Howarth}
  et~al.}{2007}]{2007MNRAS.381..433H}
{Howarth} I.~D.  et~al., 2007, \mnras, 381, 433

\bibitem[\protect\citeauthoryear{{Leutenegger}, {Cohen}, {Zsarg{\'o}},
  {Martell}, {MacArthur}, {Owocki}, {Gagn{\'e}} \& {Hillier}}{{Leutenegger}
  et~al.}{2010}]{2010ApJ...719.1767L}
{Leutenegger} M.~A.,  {Cohen} D.~H.,  {Zsarg{\'o}} J.,  {Martell} E.~M.,
  {MacArthur} J.~P.,  {Owocki} S.~P.,  {Gagn{\'e}} M.,    {Hillier} D.~J.,
  2010, \apj, 719, 1767

\bibitem[\protect\citeauthoryear{{Martins}, {Donati}, {Marcolino}, {Bouret},
  {Wade}, {Escolano} \& {Howarth}}{{Martins}
  et~al.}{2010}]{2010MNRAS.407.1423M}
{Martins} F.,  {Donati} J.,  {Marcolino} W.~L.~F.,  {Bouret} J.,  {Wade} G.~A.,
   {Escolano} C.,    {Howarth} I.~D.,  2010, \mnras, 407, 1423

\bibitem[\protect\citeauthoryear{{Mathys}, {Hubrig}, {Landstreet}, {Lanz} \&
  {Manfroid}}{{Mathys} et~al.}{1997}]{1997A&AS..123..353M}
{Mathys} G.,  {Hubrig} S.,  {Landstreet} J.~D.,  {Lanz} T.,    {Manfroid} J.,
  1997, \aaps, 123, 353

\bibitem[\protect\citeauthoryear{{Morel} et~al.,}{{Morel}
  et~al.}{2014}]{2014Msngr.157...27M}
{Morel} T.  et~al., 2014, The Messenger, 157, 27

\bibitem[\protect\citeauthoryear{{Naz{\'e}}, {Petit}, {Rinbrand}, {Cohen},
  {Owocki}, {ud-Doula} \& {Wade}}{{Naz{\'e}}
  et~al.}{2014}]{2014ApJS..215...10N}
{Naz{\'e}} Y.,  {Petit} V.,  {Rinbrand} M.,  {Cohen} D.,  {Owocki} S.,
  {ud-Doula} A.,    {Wade} G.~A.,  2014, \apjs, 215, 10

\bibitem[\protect\citeauthoryear{{Petit} et~al.,}{{Petit}
  et~al.}{2013}]{2013MNRAS.429..398P}
{Petit} V.  et~al., 2013, \mnras, 429, 398

\bibitem[\protect\citeauthoryear{{Smith}, {Brickhouse}, {Liedahl} \&
  {Raymond}}{{Smith} et~al.}{2001}]{2001ApJ...556L..91S}
{Smith} R.~K.,  {Brickhouse} N.~S.,  {Liedahl} D.~A.,    {Raymond} J.~C.,
  2001, \apjl, 556, L91

\bibitem[\protect\citeauthoryear{{Stahl} et~al.,}{{Stahl}
  et~al.}{1996}]{1996A&A...312..539S}
{Stahl} O.  et~al., 1996, \aap, 312, 539

\bibitem[\protect\citeauthoryear{{Sundqvist}, {Petit}, {Owocki}, {Wade}, {Puls}
  \& {MiMeS Collaboration}}{{Sundqvist} et~al.}{2013}]{2013MNRAS.433.2497S}
{Sundqvist} J.~O.,  {Petit} V.,  {Owocki} S.~P.,  {Wade} G.~A.,  {Puls} J.,
  {MiMeS Collaboration} 2013, \mnras, 433, 2497

\bibitem[\protect\citeauthoryear{{Sundqvist}, {ud-Doula}, {Owocki}, {Townsend},
  {Howarth} \& {Wade}}{{Sundqvist} et~al.}{2012}]{2012MNRAS.423L..21S}
{Sundqvist} J.~O.,  {ud-Doula} A.,  {Owocki} S.~P.,  {Townsend} R.~H.~D.,
  {Howarth} I.~D.,    {Wade} G.~A.,  2012, \mnras, 423, L21

\bibitem[\protect\citeauthoryear{{Townsend}, {Owocki} \& {ud-Doula}}{{Townsend}
  et~al.}{2007}]{2007MNRAS.382..139T}
{Townsend} R.~H.~D.,  {Owocki} S.~P.,    {ud-Doula} A.,  2007, \mnras, 382, 139

\bibitem[\protect\citeauthoryear{{Townsend} et~al.,}{{Townsend}
  et~al.}{2013}]{2013ApJ...769...33T}
{Townsend} R.~H.~D.  et~al., 2013, \apj, 769, 33

\bibitem[\protect\citeauthoryear{{ud-Doula}, {Owocki}, {Townsend}, {Petit} \&
  {Cohen}}{{ud-Doula} et~al.}{2014}]{2014MNRAS.441.3600U}
{ud-Doula} A.,  {Owocki} S.,  {Townsend} R.,  {Petit} V.,    {Cohen} D.,  2014,
  \mnras, 441, 3600

\bibitem[\protect\citeauthoryear{{ud-Doula}, {Owocki} \& {Townsend}}{{ud-Doula}
  et~al.}{2008}]{2008MNRAS.385...97U}
{ud-Doula} A.,  {Owocki} S.~P.,    {Townsend} R.~H.~D.,  2008, \mnras, 385, 97

\bibitem[\protect\citeauthoryear{{ud-Doula}, {Owocki} \& {Townsend}}{{ud-Doula}
  et~al.}{2009}]{2009MNRAS.392.1022U}
{ud-Doula} A.,  {Owocki} S.~P.,    {Townsend} R.~H.~D.,  2009, \mnras, 392,
  1022

\bibitem[\protect\citeauthoryear{{ud-Doula}, {Sundqvist}, {Owocki}, {Petit} \&
  {Townsend}}{{ud-Doula} et~al.}{2013}]{2013MNRAS.428.2723U}
{ud-Doula} A.,  {Sundqvist} J.~O.,  {Owocki} S.~P.,  {Petit} V.,    {Townsend}
  R.~H.~D.,  2013, \mnras, 428, 2723

\bibitem[\protect\citeauthoryear{{Verner} \& {Yakovlev}}{{Verner} \&
  {Yakovlev}}{1995}]{1995A&AS..109..125V}
{Verner} D.~A.,  {Yakovlev} D.~G.,  1995, \aaps, 109, 125

\bibitem[\protect\citeauthoryear{{Vink}, {de Koter} \& {Lamers}}{{Vink}
  et~al.}{2001}]{2001A&A...369..574V}
{Vink} J.~S.,  {de Koter} A.,    {Lamers} H.~J.~G.~L.~M.,  2001, \aap, 369, 574

\bibitem[\protect\citeauthoryear{{Wade} et~al.,}{{Wade}
  et~al.}{2011}]{2010arXiv1009.3563W}
{Wade} G.~A.  et~al., 2011, in "{C.~Neiner, G.~Wade, G.~Meynet, \& G.~Peters}"
  ed.,  IAU Symposium Vol. 272, Active OB stars. p.~118

\bibitem[\protect\citeauthoryear{{Wade} et~al.,}{{Wade}
  et~al.}{2012a}]{2012MNRAS.425.1278W}
{Wade} G.~A.  et~al., 2012a, \mnras, 425, 1278

\bibitem[\protect\citeauthoryear{{Wade} et~al.,}{{Wade}
  et~al.}{2012b}]{2012MNRAS.419.2459W}
{Wade} G.~A.  et~al., 2012b, \mnras, 419, 2459

\bibitem[\protect\citeauthoryear{{Walborn}, {Sota}, {Ma{\'{\i}}z
  Apell{\'a}niz}, {Alfaro}, {Morrell}, {Barb{\'a}}, {Arias} \&
  {Gamen}}{{Walborn} et~al.}{2010}]{2010ApJ...711L.143W}
{Walborn} N.~R.,  {Sota} A.,  {Ma{\'{\i}}z Apell{\'a}niz} J.,  {Alfaro} E.~J.,
  {Morrell} N.~I.,  {Barb{\'a}} R.~H.,  {Arias} J.~I.,    {Gamen} R.~C.,  2010,
  \apjl, 711, L143

\bibitem[\protect\citeauthoryear{{Wilms}, {Allen} \& {McCray}}{{Wilms}
  et~al.}{2000}]{2000ApJ...542..914W}
{Wilms} J.,  {Allen} A.,    {McCray} R.,  2000, \apj, 542, 914

\bibitem[\protect\citeauthoryear{{Wolk}, {Harnden} Jr., {Flaccomio}, {Micela},
  {Favata}, {Shang} \& {Feigelson}}{{Wolk} et~al.}{2005}]{2005ApJS..160..423W}
{Wolk} S.~J.,  {Harnden} Jr. F.~R.,  {Flaccomio} E.,  {Micela} G.,  {Favata}
  F.,  {Shang} H.,    {Feigelson} E.~D.,  2005, \apjs, 160, 423

\end{thebibliography}
